\documentclass[aps,prb,showpacs,twocolumn,floatfix,letterpaper,%
superscriptaddress,10pt]{revtex4-1}

\usepackage[intlimits,sumlimits]{amsmath}
\usepackage{amsfonts,amssymb}
\usepackage{bm}
\usepackage{graphicx}
\usepackage{hyperref}

\usepackage{hyperref}
\hypersetup{%
pdftitle={``Hamiltonian optics of hyperbolic polaritons in nanogranules''},%
pdfauthor={Z. Sun et al.},%
pdfpagemode={UseNone},%
pdfstartview={FitH},%
breaklinks=true,%
citecolor=blue,%
colorlinks=true,%
linkcolor=blue,%
urlcolor=blue}

\graphicspath{{Image/}}

\begin{document}

\author{Zhiyuan Sun}
\affiliation{Department of Physics, University of California San Diego, 9500 Gilman Drive, La Jolla, California 92093, USA}

\author{\'{A}. Guti\'{e}rrez-Rubio}
\affiliation{Instituto de Ciencia de Materiales de Madrid, CSIC, Cantoblanco, E-28049 Madrid, Spain}

\author{D. N. Basov}
\affiliation{Department of Physics, University of California San Diego, 9500 Gilman Drive, La Jolla, California 92093, USA}

\author{M. M. Fogler}
\email{mfogler@ucsd.edu}
\affiliation{Department of Physics, University of California San Diego, 9500 Gilman Drive, La Jolla, California 92093, USA}

\title{Hamiltonian optics of hyperbolic polaritons in nanogranules}

\begin{abstract}
Semiclassical quantization rules
and numerical calculations are applied to study polariton modes of materials
whose permittivity tensor has principal values of opposite sign (so-called hyperbolic materials).
The spectra of volume- and surface-confined polaritons are
computed for spheroidal nanogranules of hexagonal boron nitride, a natural hyperbolic crystal.
The field distribution created by polaritons excited by an external dipole source is predicted to exhibit ray-like patterns due to classical periodic orbits.
Near-field infrared imaging and Purcell-factor measurements are suggested to test these predictions.
\end{abstract}

\maketitle

\textbf{Introduction.} Recently much interest has been attracted to a class of uniaxial materials whose axial $\varepsilon_z$ and tangential $\varepsilon_{\perp}$ permittivities have opposite signs.
These hyperbolic materials (HM) possess extraordinary rays with unusual properties.
In this Letter we focus on polar dielectric HM~\cite{Fonoberov2005, Dai2014, Caldwell2014, Jacob2014,Leonid2012} where the extraordinary rays are phonon-polariton collective modes.
Our results may also apply to other HM, including
ferromagnets~\cite{Walker1957}, magnetized plasmas~\cite{Fisher1969}, artificial metamaterials~\cite{Poddubny2013}, layered superconductors~\cite{Stinson2014, Alpeggiani2013}, and liquid crystals~\cite{Pawlik2014}.

\begin{figure}
	\includegraphics[width=3.2in]{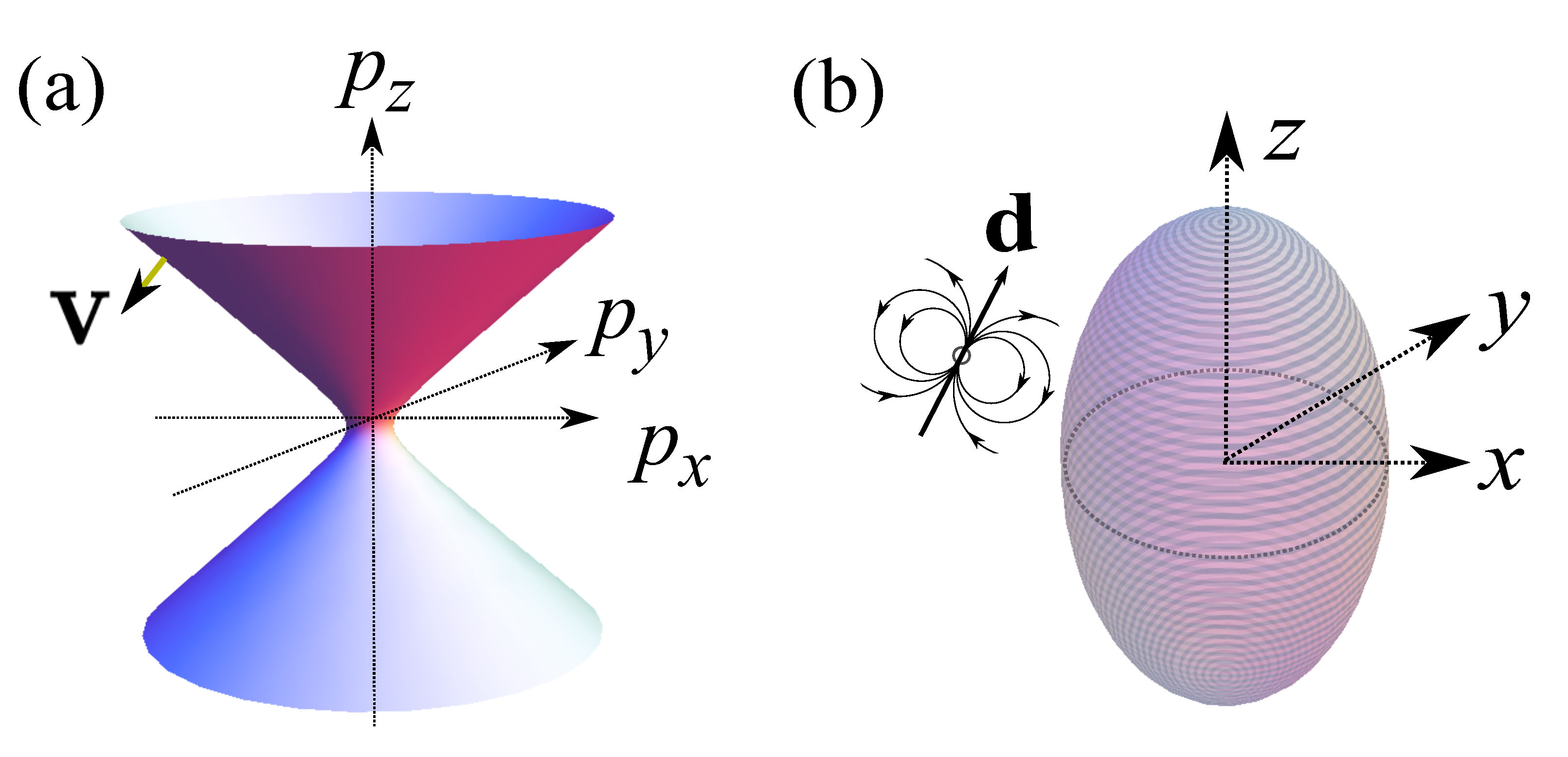}
	\caption{(a) A schematic of a polariton isofrequency surface and the group velocity $\mathbf{v}$ in a HM. The example shown is for the case $\varepsilon_{\perp} < 0$, $\varepsilon_z > 0$.
		(b) The geometry of the model studied.
		Vector $\mathbf{d}$ symbolizes an external dipole source.
	}
	\label{fig:model}
\end{figure}

The basic properties of hyperbolic polaritons are as follows.
Their isofrequency surfaces $\omega(\mathbf{p}) = \mathrm{const}$ in momentum space $\mathbf{p} = (p_x, p_y, p_z)$ are hyperboloids.
In a broad range of $|\mathbf{p}|$ from the free-space photon momentum $\omega / c$ to
an upper cutoff imposed by microscopic structure,
these hyperboloids can be approximated by cones [Figure~\ref{fig:model}(a)]
\begin{equation}
	H_B(\mathbf{p}, \omega) \equiv \varepsilon_z(\omega) p_z^2  + \varepsilon_{\perp}(\omega) (p_x^2 + p_y^2)
	=0\,.
	\label{eqn:Hamiltonian}
\end{equation}
The group velocity $\mathbf{v}(\mathbf{p}) = \partial_{\mathbf{p}}\, \omega$ is always
orthogonal to the isofrequency surface.
Hence, within the conical approximation it has a fixed angle $\alpha = \tan^{-1} \left(i
\frac{\sqrt{\varepsilon_{\perp}}}
     {\sqrt{\varepsilon_z}}
\,\right)$ with respect to the optical axis.
Such a strictly directional propagation of polaritons
may be used for sub-diffractional focusing~\cite{Li2015hpp, Dai2015a} and super-resolution imaging
known as `hyperlensing'~\cite{Poddubny2013, zhaowei.liu.2007, Jacob:06, Salandrino2006}.
Since the high-momenta polaritons remain immune to evanescent decay,
volume-confinement of polaritons inside nanogranules \cite{Caldwell2014, Xu2014, Yang2012} is possible.
Several experimental observations of such modes in
hexagonal boron nitride (hBN), a natural mid-infrared HM,
have been reported~\cite{Dai2014, Xu2014, Caldwell2014, Shi2015aap, Li2015hpp, Dai2015a}.
(This layered insulator is also known to be a premier substrate~\cite{Dean2010} or a spacer for van der Waals heterostructures~\cite{Geim2013,Fogler2014}.)
In the far-field spectroscopy~\cite{Caldwell2014} the polariton modes of hBN nanogranules show up as discrete resonances.
Remarkably, the spectrum of such resonances was found to depend primarily on the aspect ratio of the granules rather than their size or precise shape.
Exact solutions~\cite{Walker1957, Fonoberov2005, Alpeggiani2013} for spheroidal or spherical shapes enable one to compute such spectra but they do not elucidate the underlying physical picture.

In this Letter, we further develop an alternative ray optics method \cite{Jacob:07} that
makes connection to the Einstein-Brillouin-Keller (EBK) quantization~\cite{Keller1960, Gutzwiller.1990} of a classical particle inside a cavity having the same shape as the granule.
The indefinite permittivity tensor of the HM maps on the indefinite Hamiltonian $H_B(\mathbf{p}, \omega)$ of the particle, eq~\eqref{eqn:Hamiltonian}.
The EBK quantization rules are valid provided the classical motion is regular~\cite{Gutzwiller.1990}.
However, they give valuable physical insights even when it is weakly chaotic~\cite{Noeckel1996clt, Narimanov2009} or pseudointegrable~\cite{Wiersig2003hdr}.
Classical motion in a spheroidal cavity is completely integrable,
and so the EBK rules apply directly to spheroidal granules.

Nanoscale spatial distribution of the electric field produced by the polariton modes can be measured by scanning near-field optical microscopy~\cite{Dai2014, Xu2014, Dai2015a}.
To model such experiments we compute the response of a spheroidal nanogranule to an oscillating electric dipole.
We calculate the field reflected back to the dipole and the field distribution  it induces in the interior and on the surface of the spheroid. Both of them exhibit striking
geometrical patterns that correspond to periodic orbits of polaritonic rays.

\begin{figure}
	\includegraphics[width=3.2in]{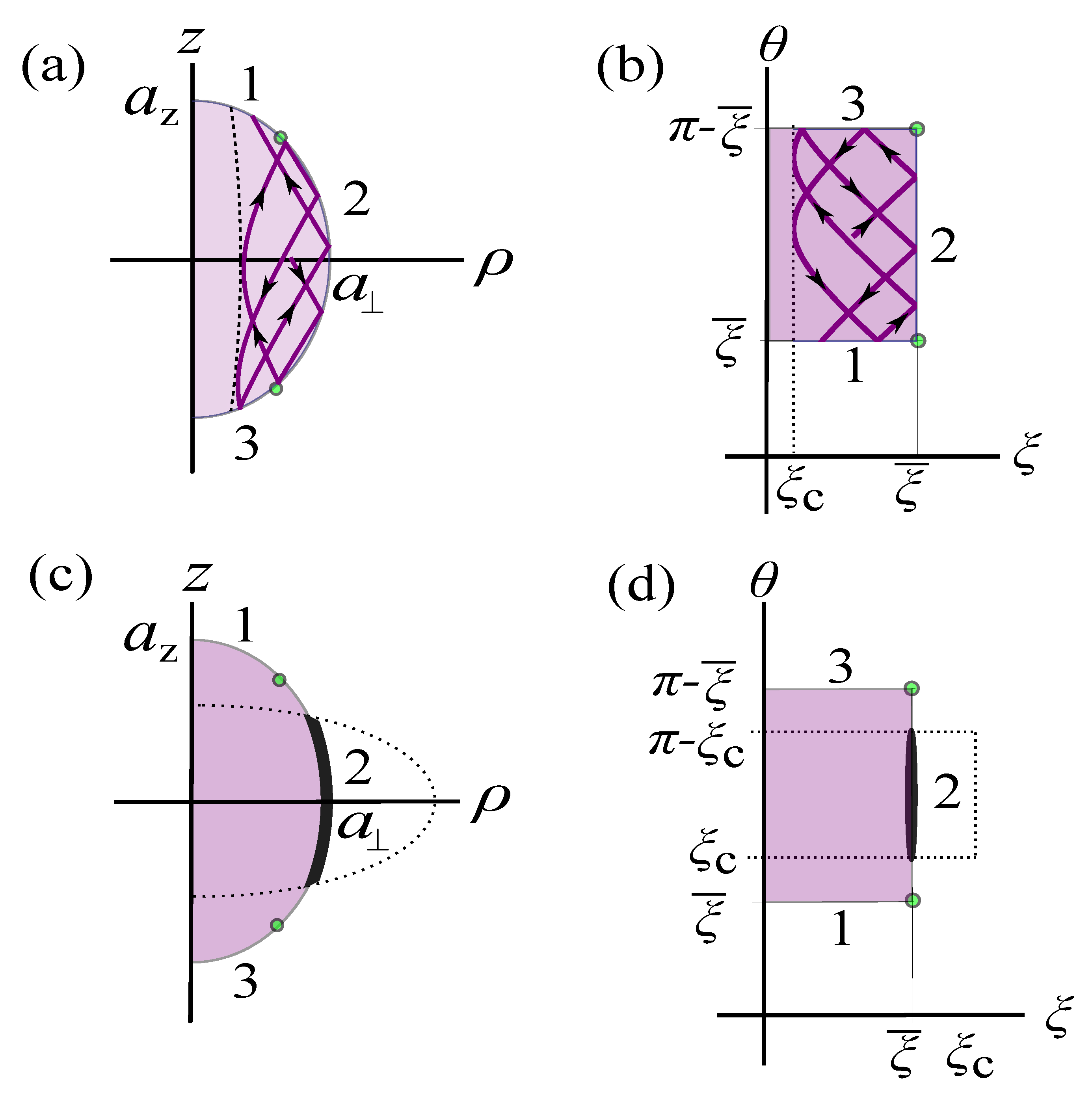}
	\caption{The correspondence of the $(\rho, z)$ and $(\xi, \theta)$ coordinate systems for the spheroid.
		$1$, $2$, $3$ label the three boundary regions.
		The green dots  separating these regions mark the points where the spheroid surface is tangent to the polariton group velocity $\mathbf{v}$.
		(a,b) the classically accessible bulk region for a wave with a caustic $\xi = \xi_c$ (dotted line) inside the spheroid. Lines inside the region are trajectories of wave packets, which are straight lines in real space although they appear as curves in our choice of coordinates. 
		(c, d) The classically accessible boundary region (thick dark line)
		for a surface wave. The caustic (dotted line) extends outside the spheroid.
	}
	\label{fig:spheroidal_coordinate}
\end{figure}

\textbf{The exact eigenmodes.}
Consider a granule that has a shape of a spheroid with the symmetry axis parallel to the optical axis ($z$-axis) of the permittivity tensor (Figure~\ref{fig:model}b).
We assume the spheroid is prolate, $a_z > a_\perp$.
(Oblate spheroids can be treated in a similar manner.) 
The cross-section of the granule in the cylindrical coordinates $\rho \equiv \sqrt{x^2 + y^2}$ and $z$ is shown in Figure~\ref{fig:spheroidal_coordinate}a,c.
We define two other sets of coordinates.
Outside the spheroid, we use the usual spheroidal ones:
\begin{equation}
	\rho = a  \sinh {\eta} \sin\theta\,, \quad
	z = a  \cosh {\eta} \cos\theta\,,
	\label{eqn:outside_coord}
\end{equation}
where ${\eta} > \bar{\eta}$, $0<\theta<\pi$, $a^2 = a_z^2 - a_{\perp}^2$,
and
\begin{equation}
	\tanh \bar{\eta} = \frac{a_{\perp}}{a_z}\,.
	\label{eqn:eta_bar}
\end{equation}
These coordinates are orthogonal and real.
Inside, Figure \ref{fig:spheroidal_coordinate}b and d, we use
\begin{equation}
	\rho = -i b \sqrt{\varepsilon_{\perp}}\,  \sin\xi \sin\theta\,, \quad
	z = b \sqrt{\varepsilon_{z}}\, \cos\xi \cos\theta\,,
	\label{eqn:inside_coord}
\end{equation}
where $0 < \xi < {\overline{\xi}}$, ${\overline{\xi}} < \theta < \pi - \overline{\xi}$, and
\begin{equation}
	\tan {\overline{\xi}} = i\, \frac{a_{\perp}}{a_z} \frac{\sqrt{\varepsilon_z}}{\sqrt{\varepsilon_{\perp}}}\,.
	\label{eqn:xi_bar}
\end{equation}
Parameter
$b = (\varepsilon_z^{-1} a_z^2 - \varepsilon_{\perp}^{-1} a_{\perp}^2)^{1/2}$
is real if $\varepsilon_\perp < 0$, $\varepsilon_z > 0$, and imaginary if both the signs are reversed.
The coordinates $(\xi, \theta)$ are real and leave the permittivity tensor diagonal.
However, they are nonorthogonal.

We assume that granule is suspended in vacuum and
that its size is much smaller than $c / \omega$.
In this case the quasi-static approximation for the electric field is valid,
$\mathbf{E} = -\partial_{\mathbf{r}} \Phi$.
The scalar potential $\Phi$ is represented by two different functions
inside and outside the particle: $\Phi_1$ and $\Phi_2$.
The inner potential
obeys the Walker equation~\cite{Walker1957}
\begin{equation}
	[\varepsilon_z \partial_z^2  + \varepsilon_{\perp} (\partial_x^2 + \partial_y^2)] \Phi_1 = 0\,.
	\label{eqn:Walker}
\end{equation}
The outer potential $\Phi_2$ satisfies the Laplace equation.
The potential and the normal component of the displacement must be continuous across the spheroid surface.
In the chosen coordinates, these boundary conditions become separable,
which enables one to find the analytical solutions~\cite{Fonoberov2005, Walker1957} of this eigenproblem:
\begin{subequations}
	\label{eqn:Phi_1_and_2}
	\begin{align}
		\Phi_1 &\propto \mathsf{P}^m_l(\cos\xi) \mathsf{P}^m_l(\cos\theta) {e}^{{i} m \phi}\,,   &&  0 < \xi < {\overline{\xi}}\,,  \label{eqn:solution_inside}
		\\
		\Phi_2 &\propto Q^m_l(\cosh\eta) \mathsf{P}^m_l(\cos\theta) {e}^{{i} m \phi}\,,   && {\eta} > \bar{\eta}\,.
		\label{eqn:solution_outside}
	\end{align}
\end{subequations}
Here $m$, $l$ are integers,
$\mathsf{P}^m_l(z)$, $Q^m_l(z)$ are
the associated Legendre functions of the first and the second kinds,
respectively,
and $\phi$ is the polar coordinate in the $x$--$y$ plane.
The boundary conditions are satisfied provided~\cite{Walker1957, Fonoberov2005}
\begin{equation}
	i \sqrt{\varepsilon_{\perp}} \sqrt{\varepsilon_z}\, \frac{d}{d \overline{\xi}} \,
	\ln \mathsf{P}^{m}_{l}(\cos{\overline{\xi}})
	= \frac{d}{d \bar{\eta}}\, \ln Q^{m}_{l}(\cosh \bar{\eta})\,.
	\label{eqn:exact}
\end{equation}
For each $m$ and $l$ this equation gives us several solutions for the eigenfrequency $\omega$ (contained implicitly in $\varepsilon_\perp$, $\varepsilon_z$) that can be indexed by another integer $n$.
The total number of such solutions is equal to $l$ for $m = 0$ and $l - |m| + 1$ for $m \neq 0$ (Supporting information).
Note that eq~\eqref{eqn:exact} depends only on the aspect ratio and not the size of the spheroid.
This is consistent with the scale-invariance of eq \eqref{eqn:Walker}.
However, the physical picture is not clear from this exact solution.
Next, we present an alternative derivation in terms of a more intuitive ray-optics approach.


\textbf{Hamiltonian optics.}
 Ray or geometrical optics is a well established
approach to study propagation of light on
scales longer than the photon wavelength. HMs are a new arena for ray optics in which photons are replaced by excitations of much shorter wavelength --- polaritons --- in the case of hyperbolic polar insulator.
This approach has been previously applied to HM of cylindrical geometry \cite{Jacob:07, Narimanov2009}.
Here we study a spheroidal granule and address both the ray and the wave optics effects within the quasi-static approximation.
The derivation of the ray picture
starts with seeking the scalar potential in the form
\begin{equation}
	\Phi_1(\mathbf{r}) = \sum_j A_j(\mathbf{r}) {e}^{{i} S_j(\mathbf{r})}
	\,,
	\label{eqn:eikonal}
\end{equation}
where the phases (or eikonals) $S_j(\mathbf{r})$ vary much faster than the amplitudes $A_j(\mathbf{r})$.
Substituting eq~\eqref{eqn:eikonal} into eq~\eqref{eqn:Walker} and keeping only the leading terms, quadratic in momenta $\mathbf{p}_j(\mathbf{r}) = \partial_{\mathbf{r}} S_j$, one obtains (for each $j$) the Hamilton-Jacobi equation~\eqref{eqn:Hamiltonian} of a fictitious classical system with the `optical'
Hamiltonian $H_B(\mathbf{p}_j, \omega)$.
The EBK quantization is possible if the number of different $j$ in eq~\eqref{eqn:eikonal} is finite.
For the spheroid four terms suffice, corresponding to
the different sign choices of the momenta.
To describe $\theta$- and $\xi$-motions one needs two terms each because our fictitious
particle can propagate in two opposite directions between the surface and the caustics.
The $\phi$-motion has no caustic and one term is enough.
We must clarify that `motion' and `propagation' 
refer to the geometry of the phase-space flow, not to the actual time evolution of coordinates and momenta.
The velocity $\mathbf{v}_b = \partial_{\mathbf{p}} H_B$ of the fictitious particle deduced from the optical Hamiltonian  is different from
the group velocity of an actual polariton
\begin{equation}
	\mathbf{v} = \frac{\partial \omega}{\partial \mathbf{p}}
	= -\left( \frac{\partial H_B}{\partial \omega} \right)_{H_B=0}^{-1}
	\mathbf{v}_b\,.
	\label{eqn:velocity}
\end{equation}
However, $\mathbf{v}_b$ and $\mathbf{v}$ are always parallel to each other.
Therefore, if a fictitious classical particle with the conserved energy $H_B(\mathbf{p}, \omega) = 0$
and a polariton wavepacket of frequency $\omega$ are launched at the same initial point $(\mathbf{p}, \mathbf{r})$,
the geometrical shape of their phase-space trajectories will
be identical.
This identity is well known in the Hamiltonian formulation of geometrical optics~\cite{Chaves.2008}.
Here we adopt it for hyperbolic polaritons.
The EBK quantization rules~\cite{Keller1958, Keller1960} can also be directly adopted for our problem because they are formulated in terms of contour integrals in the phase-space.
Therefore, to compute polariton eigenmodes of an arbitrary nanostructure made of HM, we need to quantize the motion of a single particle bouncing inside a cavity of the same shape.

Two unusual circumstances still have to be handled.
First, the Hamiltonian of our fictitious particle is indefinite.
Second, the reflection rule and the corresponding phase shift at the surface are determined by
the boundary conditions.
For spheroidal nanogranule (Figure~\ref{fig:model}b),
both of these peculiarities prove to be tractable in the coordinate system defined above.
Performing the canonical transformation to the new momenta $p_\xi, p_\theta$, we find
\begin{equation}
	H_B = \frac{p_{\xi}^2 - p_{\theta}^2}{\sin^2\xi - \sin^2\theta}
	- \frac{p_{\phi}^2}{\sin^2\xi \sin^2\theta}\,.
	\label{eqn:Hamiltonian_inside}
\end{equation}
%
The classical motion governed by Hamiltonian eq~\eqref{eqn:Hamiltonian_inside} is separable and so integrable.
(Unfortunately, in the class of smooth convex shapes, only ellipsoids and spheroids as their particular case appear to be integrable~\cite{Amiran1997isp}.)

The EBK quantization rules are in the form
\begin{subequations}
	\label{eqn:EBK}
	\begin{align}
		\oint p_{\xi} d\xi -\frac{\pi}{2} + \delta &=  2\pi \nu \,,
		\label{eqn:EBK_xi}\\
		\oint p_{\theta} d\theta + 2 \delta &= -2\pi \lambda\,,
		\label{eqn:EBK_theta}\\
		2\pi p_\phi &= 2\pi \mu\,.
		\label{eqn:EBK_phi}
	\end{align}
\end{subequations}
Here each integral is taken over a closed-loop contour in a respective coordinate (cf.~Figure~\ref{fig:spheroidal_coordinate}b). 
The phase shift $\delta$ is constant everywhere on the surface (Supporting information),
as demanded by the separable form of the exact solution, eq~\eqref{eqn:Phi_1_and_2}.
These integrals can be evaluated in terms of elementary functions.
Comparing those expressions with
the asymptotic formulas~\cite{Keller1960} for Legendre functions,
it is easy to establish
the correspondence between the EBK quantum numbers $\lambda$, $\nu$, $\mu$ and
the indices $l$, $m$, $n$ in the exact solution
(Supporting information):
\begin{equation}
	l = 2\nu + \lambda + \mu\,, \quad m = \mu\,,
	\quad n = \nu \,.
	\label{eqn:lambda}
\end{equation}
We refer to the $\nu > 0$ eigenmodes as the bulk modes and to those with $\nu = 0$ as the surface ones.
The scalar potential $\Phi_1$ of the bulk modes oscillates inside the granule along the `radial' direction $\xi$ whereas that of the surface modes monotonically increases with $\xi$ and reaches a maximum at the surface.

\begin{figure}
	\includegraphics[width=3.5in]{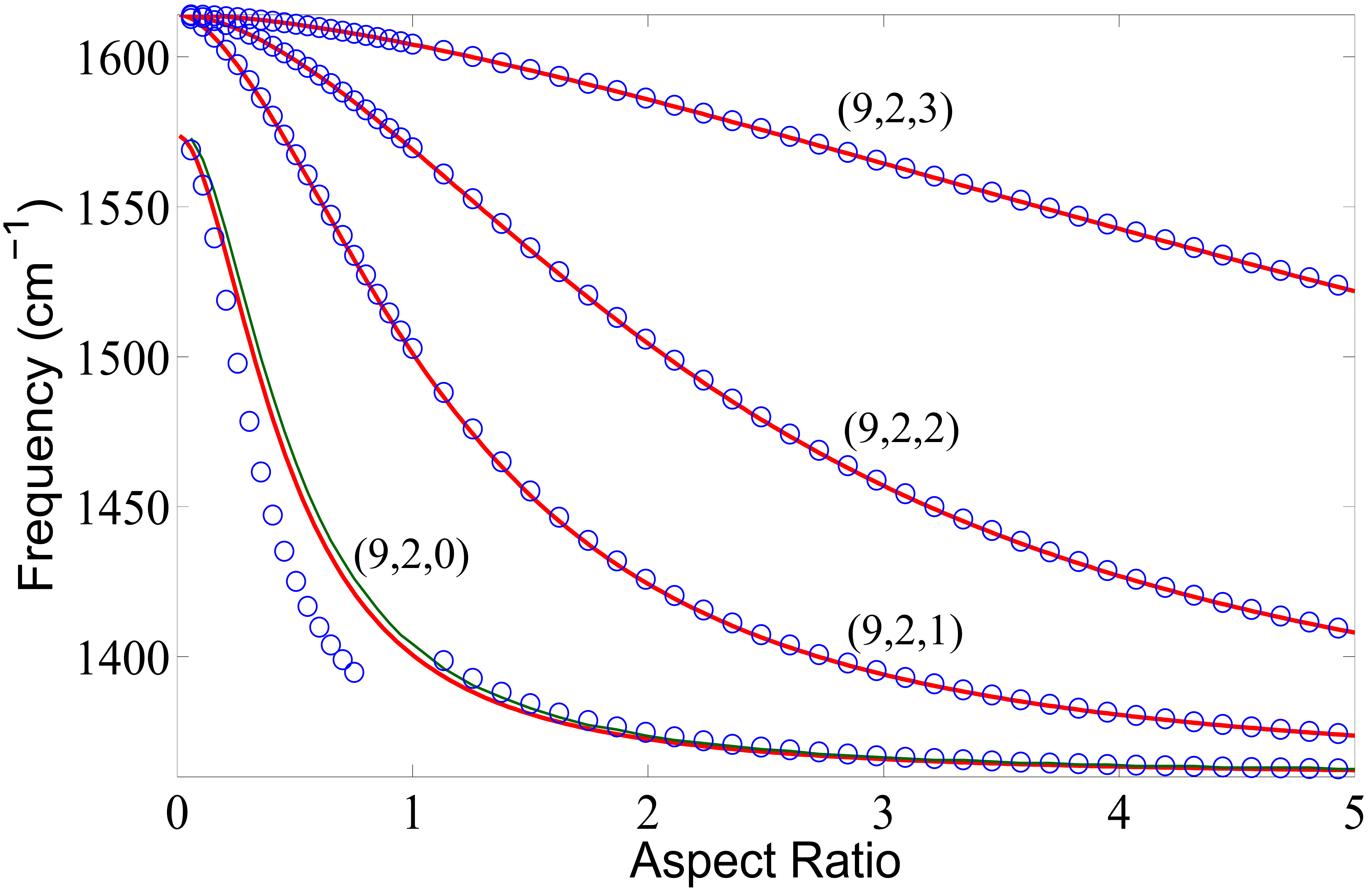}
	\caption{Eigenfrequencies of polariton modes in an hBN spheroid as functions of the aspect ratio $\mathcal{A} = {a_\perp}/{a_z}$. The red lines are exact solutions of eq~\eqref{eqn:exact}. The blue circles are from the EBK quantization method. The labels are the mode indices  $(l, m, n)$ with $n = 0$ and $n > 0$ being surface and bulk modes, respectively. For the $(9,2,0)$ branch which is classified as a surface mode, the left part of the blue circles is from the surface EBK quantization method, and the right part is from the bulk one.  The dark green line is from the uniform approximation method (Supporting information).
	}
	\label{fig:Keller_Exact}
\end{figure}

To compare the EBK results with the exact solution, we calculated the eigenmode spectra of an hBN spheroid as a function of its aspect ratio.
The measured~\cite{Caldwell2014} optical constants of hBN were used except the damping was neglected in order to obtain real solutions for $\omega$.
Examples of these calculations are shown in Figure~\ref{fig:Keller_Exact}.
The EBK is expected to be asymptotically exact at large quantum numbers but
as one can see from Figure~\ref{fig:Keller_Exact}, an excellent agreement is reached  for the bulk modes (the top three curves) already for modest $l$, $m$, and $n$.
On the other hand, the $(9, 2, 0)$ surface mode (the bottom curve) shows some deviations from the exact result at intermediate aspect ratios.
We discuss such modes in more detail below.

\textbf{Surface modes.}
The hyperbolic surface modes (HSM)~\cite{Jacob2008} are similar~\cite{Cojocaru2014} to Dyakonov surface waves~\cite{Takayama2008, Takayama2014, Pulsifer2013} of uniaxial materials with positive-definite permittivity tensor.
However, the HSMs have several new properties.
Unlike the standard Dyakonov waves,
the momenta and therefore achievable degree of confinement for the HSM are limited only by microscopic (for hBN, atomic) structure.
The HSM are robust to surface defects in the sense that there can only be three other fixed directions for the defect-scattered wave.
This is a stronger angular restriction than the absence of electron backscattering in topological insulators~\cite{Qi2011} and graphene ~\cite{CastroNeto2009}.
Finally, compared to surface plasmons in metals, which lack any directionality,
the HSM of polar insulators should exhibit a much lower damping as the they are free of electronic losses.

In the present case of the spheroid, the HSM correspond to the EBK quantum numbers $\nu = 0$ (and so to $n = 0$).
The $\xi$-coordinate of the caustic is given by
\begin{equation}
	\sin \xi_c = \frac{m}{l + \frac{1}{2}}
	= \frac{\mu}{\lambda + \mu + \frac{1}{2}}\,,
	\label{eqn:xi_c}
\end{equation}
see Supporting information.
For $l$ and $m$ fixed, $\xi_c$ is independent of the aspect ratio $\mathcal{A} = a_\perp / a_z$.
The coordinate  $\overline{\xi}$ of the spheroid surface [eq~\eqref{eqn:eta_bar}] increases with $\mathcal{A}$.
At large enough $\mathcal{A}$, we have $\overline{\xi} > \xi_c$,
see Figure~\ref{fig:spheroidal_coordinate}b.
This is similar to bulk modes (i.e., $n > 0$ modes) except the caustic is now very close to the surface.
At small $\mathcal{A}$, we have $\overline{\xi} < \xi_c$,
so the caustic extends beyond the surface, Figure~\ref{fig:spheroidal_coordinate}d.
Momentum $p_\xi$ is imaginary inside the spheroid and eq~\eqref{eqn:EBK_xi} cannot be satisfied.
[In fact, eq~\eqref{eqn:EBK_xi} fails to give a solution already shortly before
$\overline{\xi}$ drops below $\xi_c$.]
The structure of the HSM in this case can be understood from the following physical picture.
The HSM must exponentially decrease into the interior of the granule.
It can be viewed as a
wave with an imaginary $p_\xi$ outgoing from the surface into the bulk,
i.e., a surface-reflected wave generated in the absence of an incident one.
Therefore, the amplitude ratio of the two waves is formally infinite.
On the other hand, this ratio equals to $e^{i\delta}$, and so
the HSM with imaginary $p_\xi$ exists if $e^{i\delta} \to \infty$ or
$
	\tan \frac{\delta}{2} = -i\,.
	\label{eqn:HSM_condition}
$
Using this condition, eq~\eqref{eqn:xi_c}, and expression for the phase shift $\delta$ (Supporting information),
one can numerically solve for the eigenfrequency of the HSM for any given
$l$ and $m$.
The results of such calculations are illustrated in the left part of the $(9, 2, 0)$ curve in Figure~\ref{fig:Keller_Exact}.
They demonstrate a good agreement with the exact dispersion curve at small $\mathcal{A}$ where this approach is justified.
We note that the agreement can be greatly improved if the EBK formalism is replaced by the so-called uniform approximation, which also enables one to smoothly connect small and large $\mathcal{A}$ parts of the dispersion curve (Supporting information).
Lastly, one can check that the quantum numbers $\lambda = l  - m$ and $\mu = m$ of the HSM still obey the EBK rules,
\begin{equation}
	\oint p_{\theta} d\theta = -\pi (2\lambda - 1)\,,
	\quad
	p_\phi = \mu\,,
	\label{eqn:EBK_HSP}
\end{equation}
applied now to the effective \textit{surface} Hamiltonian
\begin{equation}
	H_S = p_{\theta}^2 + \left( \frac{1}{\sin^2\theta} - \frac{1}{\sin^2\xi_c} \right) p_{\phi}^2
	\label{eqn:Surface_Hamiltonian}
\end{equation}
at energy $H_S = 0$.

\begin{figure}
\includegraphics[width=3.5in]{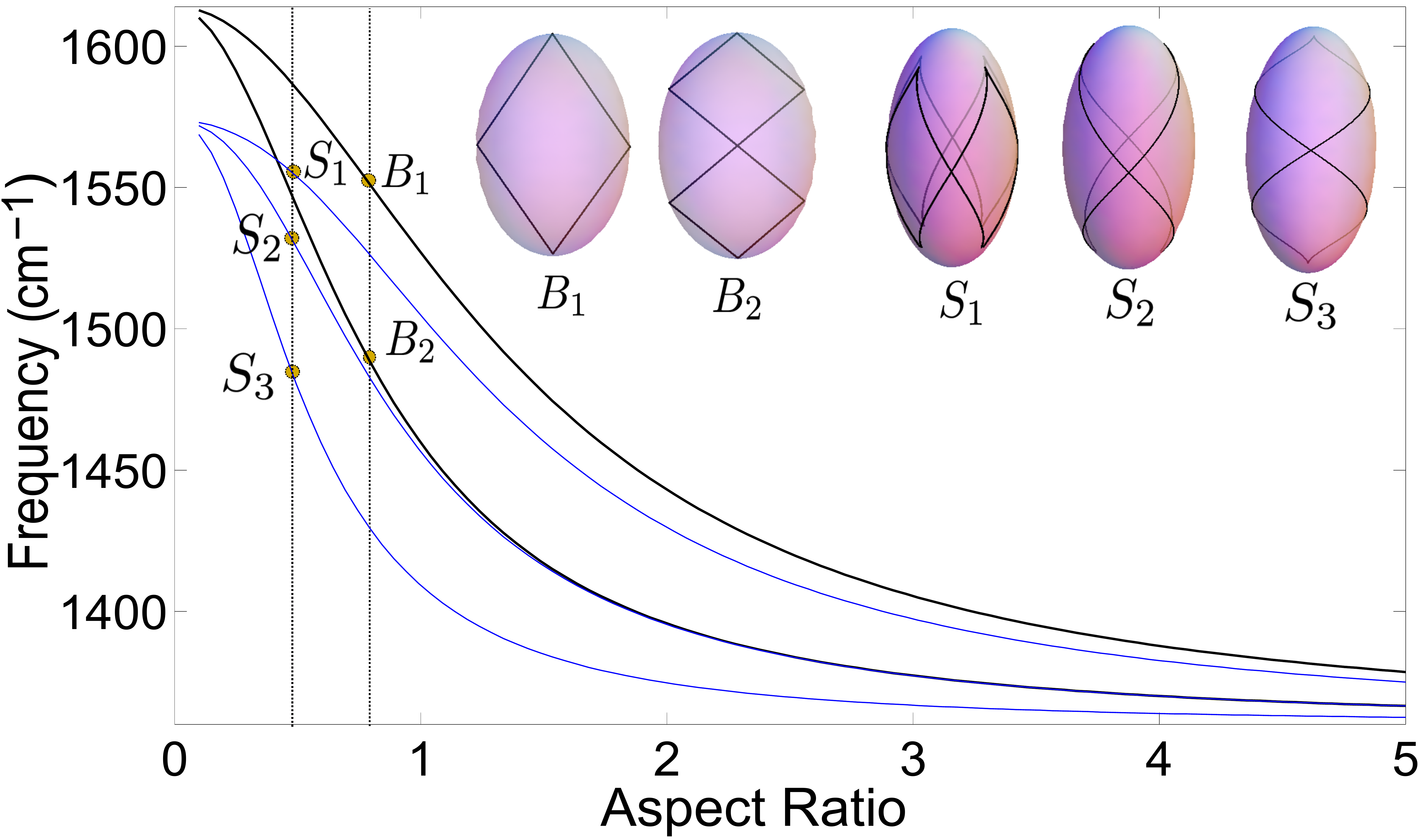}
\caption{Frequencies of representative periodic orbits as functions of the spheroid aspect ratio. The thick black lines are for the bulk orbits $B_1$ and $B_2$ with the period ratios $\tau^{-1}_\xi:\tau^{-1}_\theta:\tau^{-1}_\phi = -2:1:0$ and $-4:1:0$. The thin blue lines are for the surface orbits $S_1$, $S_2$, $S_3$ with the period ratios $\tau^{-1}_\theta: \tau^{-1}_\phi = 2:1$, $1:1$, and $1:2$. The insets show such orbits in the real space.
For the surface orbits, they include all the orbits of the given type passing through the equatorial point facing the viewer: two for each $S_1$ and $S_2$ and one for $S_3$. Despite similarity to Figure~\ref{fig:Keller_Exact}, there is no direct relation between the dispersion curves of \textit{classical} periodic orbits and those of \textit{quantized} eigenmodes, either surface or bulk ones.}
\label{fig:Periodic_Disperion}
\end{figure}

While the assumption of a spheroidal granule simplifies the theoretical analysis,
one may ask if
is it possible to make some correspondence between such a theory 
and the available experiments that were all done with HM samples of
non-spheroidal shapes.
Our tentative answer is as follows.
The modes observed in truncated hBN nanocones,~\cite{Caldwell2014}
which were previously called `volume-confined' are, in fact, similar to a subset of our HSM, specifically, $(l,m,n)=(l,0,0)$ and $(l,1,0)$ modes.
The modes of
cuboidal hyperbolic metamaterials~\cite{Yang2012}
and hBN slabs~\cite{Dai2014, Dai2015a} are conceptually similar to our bulk modes.
However, indexing them with $l$, $m$, or $n$ would be tenuous as
the conserved quantities in such systems are considerably different from those of prolate spheroids. (For example, translational momenta in slabs \textit{vs.} angular momenta in spheroids.)

\textbf{Periodic orbits.}
Classical dynamics can prominently impact the structure of quantum energies and quantum wavefunctions~\cite{Gutzwiller.1990}.
In particular, the latter may contain `scars' --- enhanced intensity lines --- along these classical trajectories.
An orbit on an invariant torus\cite{Arnold1989} defined by a set of coordinates $i$
is periodic (closed)
if the ratios of the individual periods of motion $\tau_i$ are rational numbers.
For our bulk orbits, the condition is
$\tau_\xi : \tau_\theta : \tau_\phi = z_1 : z_2 : z_3$
and for the surface periodic orbits, it is $\tau_\theta : \tau_\phi = z_1 : z_2$, where all $z_i$'s are integers.
Figure \ref{fig:Periodic_Disperion} shows the eigenfrequencies of two bulk and three surface periodic orbits as functions of the spheroid aspect ratio.
We expect that at such frequencies the field distribution created by polaritons excited by external sources should exhibit regular geometrical patterns.
Below we verify this prediction by direct numerical calculations.

\textbf{Response to a dipole.}
A peculiar property of the spheroid is that the dipole moment of all $m > 1$ modes exactly vanishes, and so they have extremely weak coupling to far-field radiation.
Furthermore, while the dipole moment of the bulk, i.e., $n > 0$ modes is nonzero, it is quite small.
Detection of all such modes
in conventional optics experiments~\cite{Caldwell2014}
will be challenging.
However, observation of these modes may be possible using scanning near-field optical microscopy.
The latter technique utilizes a sharp metalized tip to perturb and measure the system response locally.
Crudely, one can model the tip as a point dipole and the measured signal as the electric field created by the system at the location of such a dipole, see \cite{Zhang2012, McLeod2014, Jiang2015} and references therein.
The same quantity determines Purcell's factor --- the enhancement of the radiative decay of a dipole emitter \cite{Purcell.1946}.

\begin{figure}
\includegraphics[width=3.5in]{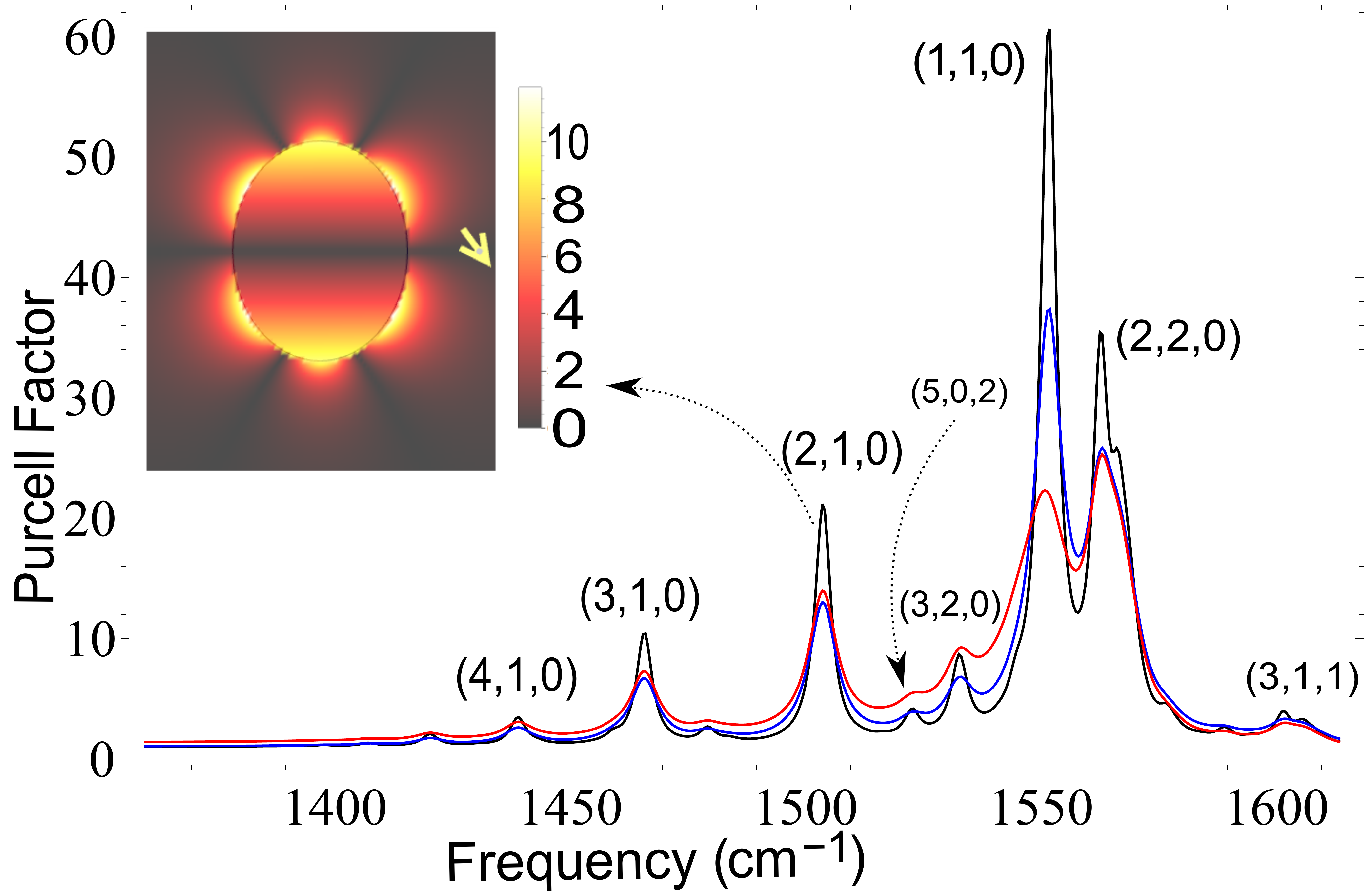}
\caption{Purcell's factor of a dipole emitter near a hBN spheroid of aspect ratio
$\mathcal{A} = \tanh 1 = 0.761$ and
long semi-axis $a_z = 500 \, \mathrm{nm}$.
The emitter is positioned on the $x$-axis (the small yellow arrow). The black and blue lines are computed assuming hBN phonon damping rate $\Gamma=4.0 \, \mathrm{cm}^{-1}$ and $\Gamma=7.0 \, \mathrm{cm}^{-1}$, respectively.
The red line is for $\Gamma=7.0 \, \mathrm{cm}^{-1}$ with
radiative damping included (see Supporting information).  
The inset depicts the distribution of the absolute value of the $E_x$ electric field component in the $x$--$z$ plane at
the $(2, 1, 0)$ resonance frequency.
}
\label{fig:purcell_mode}
\end{figure}

We assume that the emitter and its dipole moment $\mathbf{d}$ are in the $x$-$z$ plane.
(Here and below the common factor $e^{-i\omega t}$ is suppressed.)
We expand the inner and outer potentials in terms of spheroidal harmonics, i.e., the expressions appearing on the right-hand side of eqs~\eqref{eqn:solution_inside} and \eqref{eqn:solution_outside}, cf.~Supporting information for details.
In an ideal lossless HM, the electric field outside (reflected by the surface of the spheroid) would diverge at each eigenfrequency.
If the measured optical constants~\cite{Caldwell2014} of hBN are used,
this divergence is replaced by a finite-width resonance.
Purcell's factor, which is proportional to the field induced by the reflected wave at the dipole position, exhibits resonances as well,
see Figure~\ref{fig:purcell_mode}.
The strength of the resonances and the frequency spacing between them decrease as the indices $l$, $m$, and $n$ increase.
As a result, low-order resonances give distinct sharp peaks while high-order resonances merge into a smoothly varying background.
The latter is similar to the broadband
Purcell effect near the surface of an infinite HM~\cite{Poddubny2011, Poddubny2012,Jacob2012,Noginov:10,Jacob2010purcell}.
The major resonances are due to the $(l,1,0)$ modes.
Note that the perturbing dipole is assumed to have the same amplitude $\mathbf{d}$ at all $\omega$.
However, in the scanning near-field experiments instead of such a fixed dipole, one has a polarizable tip.
The back reaction of the nanogranule on the tip
is expected to cause a small but observable red shift of the resonances.
This shift can be 
modeled using recently developed analytical and numerical approaches~\cite{Zhang2012, McLeod2014, Jiang2015}
and studied experimentally by comparing
the far-field spectrum of a sparse array of identical granules~\cite{Caldwell2014} with the near-field spectrum of a single granule.
Both types of efforts can be subjects of a future work.

The electric field distribution at sharp resonances is dominated
by the resonance mode. An example is shown in the inset of Figure~\ref{fig:purcell_mode} for $(l, m, n) = (2, 1, 0)$.
This field distribution has the nodal structure of the spherical harmonic but shows no `scars'.
However, ray-like patterns do appear at the periodic orbit frequencies.
Figure~\ref{fig:field_bulk}~$B_1$ depicts the field distribution at the frequency of the bulk periodic orbit $B_1$ of Figure~\ref{fig:Periodic_Disperion}.
The shape of the high-intensity ray patterns matches
the classical periodic orbits (magenta lines).
The reason why they dominate the field distribution
can be understood by imagining that it is a superposition of fields created by wavepackets launched from a finite-size region facing the dipole.
Wavepackets whose launch points belong
to a short periodic orbit create a strongly concentrated electric field.
Other wavepackets follow quasiperiodic classical trajectories that spread all over the spheroid, giving an approximately uniform background.
Similar behavior is found near the frequency of the periodic orbit $B_2$, see Figure~\ref{fig:field_bulk}~$B_2$. 

Near-field imaging experiments are expected to be most sensitive to the electric field distribution on the surface of the granule.
Panels $S_1$--$S_3$ of Figure~\ref{fig:field_bulk} show examples of such distributions projected on the $x$--$z$ plane.
They demonstrate directional ray patterns at the frequencies of the surface periodic orbits $S_1$--$S_3$ (Figure~\ref{fig:Periodic_Disperion}).

\begin{figure}
	\includegraphics[width=3.2in]{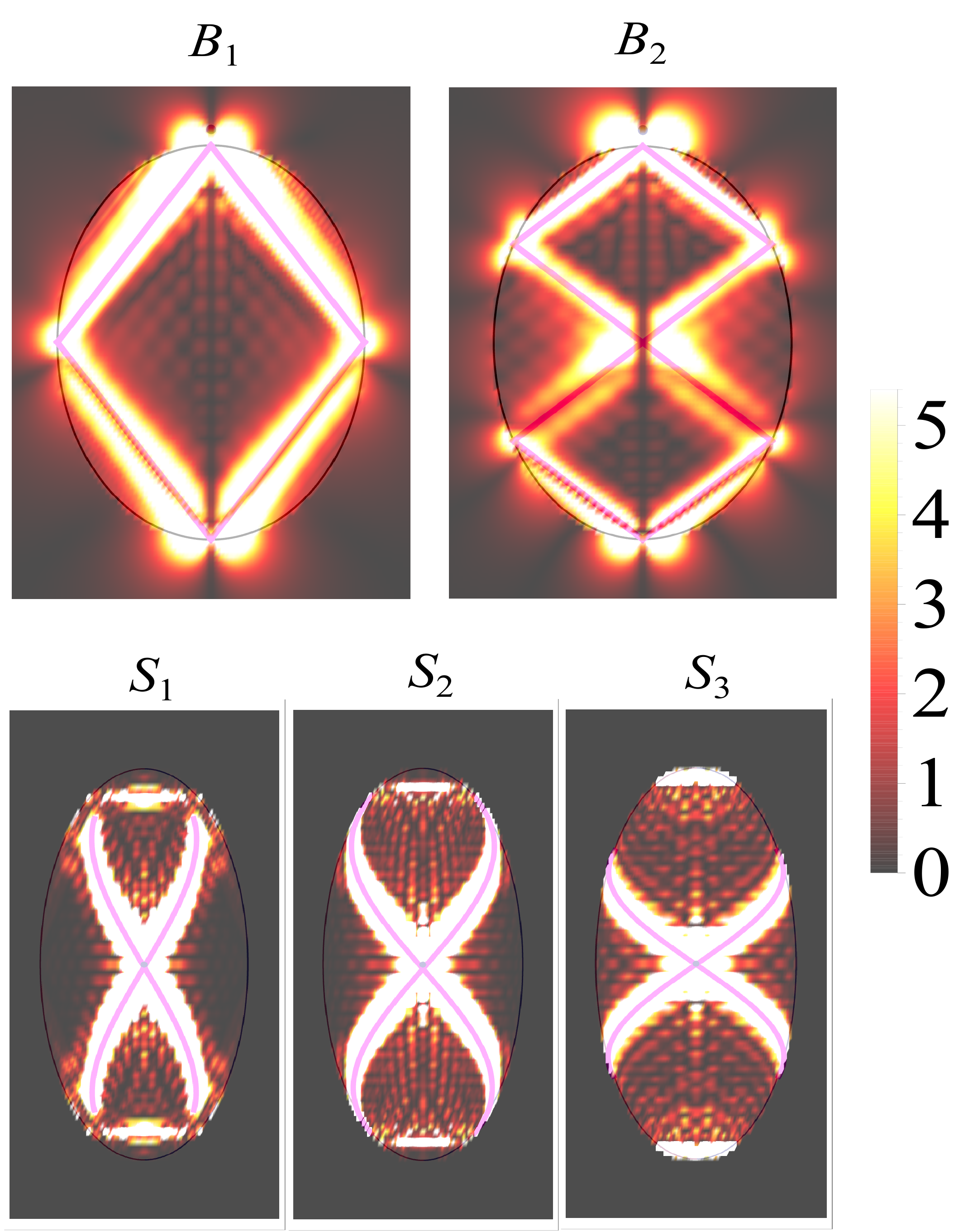}
	\caption{$B_1$, $B_2$: False color plot of $|E_x|$ in a meridional cross section of an hBN spheroid due to a vertically polarized dipole source just above the north pole. The phonon damping rate is $\Gamma=7.0 \, \mathrm{cm}^{-1}$. Outside the spheroid, the dipole's own field is subtracted away, for clarity.
		The frequencies in $B_1$, $B_2$ are $1555$, $1494 \,\text{cm}^{-1}$, which match the frequencies of $B_1$ and $B_2$ in Figure~\ref{fig:Periodic_Disperion} for the chosen aspect ratio $\mathcal{A} = \tanh 1 \approx 0.761$.
		The magenta lines are the bulk periodic orbits.
		$S_1$--$S_3$: False color plot of $E^2_z$ at the surface of the spheroid projected onto the meridional plane. The dipole is just above the surface at the center of each plot. The frequencies in $S_1$--$S_3$ are $1557$, $1535$, $1488\,\text{cm}^{-1}$, same as in Figure \ref{fig:Periodic_Disperion} for the chosen aspect ratio $\mathcal{A} = \tanh 0.5 = 0.462 $. The magenta lines are the surface periodic orbits.}
	\label{fig:field_bulk}
\end{figure}

\textbf{Conclusions.}
We investigated basic properties of confined polariton modes in spheroidal nanogranules of polar hyperbolic materials.
A physically transparent ray-optics method for computing eigenfrequencies and wavepacket dynamics of the polaritons was presented and
its accuracy verified by comparison with the exact analytical results and numerical simulations.
We also suggested how to probe these polariton modes experimentally using external dipole sources and/or scanned near-field optical microscopy.

There is a number of other directions to explore.
For example, we restricted our analysis to the hyperbolic regime
realized at frequencies inside the hBN Reststrahlen bands.
The change of the polariton isofrequency surfaces from hyperbolic to elliptical at the extremes of these bands is a topological transition.
One may want to investigate signatures of this intriguing transition
in, e.g., Purcell's factor.~\cite{Krishnamoorthy2012}

The studied phenomena may also have far-reaching technological implications.
One can imagine a whole new class of polaritonic devices that would include nanoresonantors, hyperlenses, infrared photon sources, \textit{etc}.
Such devices would be deeply sub-diffractional and low-loss because phonon-polaritons are immune to electronic losses that plague conventional metal-based plasmonics.
Our general approach may be useful for design and optimization of these devices.

We thank A. Rela\~no and M. Berry for useful discussions and also F.~Guinea for comments on the manuscript.

\smallskip
\noindent$\blacksquare$ AUTHOR INFORMATION

\noindent{}Corresponding Author:

\noindent{}$^*$E-mail: mfogler@ucsd.edu

\smallskip
\noindent{}Author Contributions

\noindent{}Z.S. and A.-G.R. carried out theoretical calculations.
D.N.B. and M.M.F. supervised and guided the work.
The manuscript was written through contributions from all
authors.
All authors have given approval to the final version of the
manuscript.

\smallskip
\noindent{}\textbf{Funding}

\noindent{}The work at UCSD is supported by
the University of California Office of the President and 
by the DOE-BES Grant DE-SC00122592.
A.~G.-R. is supported by MINECO (Spain) through Grant No.~FIS2011-23713 and by the European Research Council Advanced
Grant (Contract No.~290846).

\smallskip
\noindent{}\textbf{Notes}

\noindent{}The authors declare no competing financial interest.

\newpage
\centerline{\Large{\textbf{Supporting information}}}

\setcounter{equation}{0}
\renewcommand{\theequation}{{S}\arabic{equation}}

\setcounter{figure}{0}
\renewcommand{\figurename}{\textbf{Figure}}
\renewcommand{\thefigure}{{S}\arabic{figure}}

\newcommand{\unit}[1]{\,\mathrm{#1}} 

\section{Eigenmode dispersion}
\label{sec:eigenmode}

\subsection{Radial index of the modes}

A nanogranule made of a hyperbolic material possesses multiple bulk polariton modes corresponding to the same `angular' indices $l$ and $m$.
Such modes can be indexed with the `radial' quantum number $n$, as described below.
Consider the exact eigenmode equation [eq~(9) of the main text]:
\begin{equation}
i \sqrt{\varepsilon_{\perp}} \sqrt{\varepsilon_z} \, \frac{d}{d \overline{\xi}} \,
  \ln \mathsf{P}^{m}_{l}(\cos{\overline{\xi}})
  = \frac{d}{d \bar{\eta}}\, \ln Q^{m}_{l}(\cosh \bar{\eta})\,.
\label{eqn:exact}
\end{equation}
Following Walker~\cite{Walker1957}, this equation can be written as
\begin{equation}
\begin{split}
\varepsilon_z \left( \frac{|m|}{\tan^2 \overline{\xi}}  + k
+ \sum\limits_{i \,=\, 1}^{N}
 \frac{2}{ x^2_i \tan^2 \overline{\xi} - 1 + x^2_i} \right)&\\
= - \cosh\bar{\eta} \frac{d}{d \cosh\bar{\eta}}  \ln Q^{m}_{l}(\cosh \bar{\eta})\,,&
\end{split}
\label{eqn:exact_II}
\end{equation}
where $0 < x_i < 1$ are the positive roots of the Legendre function $\mathsf{P}^{m}_{l}(x)$ sorted in the ascending order,
$N = [(l - |m|) / 2]$ is the number of such roots, $[z]$ is the integer part of $z$, $k \equiv 2 N - n + |m|$ is either $0$ or $-1$, and
$\tan \overline{\xi}$ is defined by eq~(5) of the main text:
\begin{equation}
\tan {\overline{\xi}} = i \frac{a_{\perp}}{a_z} \frac{\sqrt{\varepsilon_z}}{\sqrt{\varepsilon_{\perp}}}\,.
\label{eqn:xi_bar}
\end{equation}
The right-hand side of eq~\eqref{eqn:exact_II} is a positive finite number,
while the left-hand side is a sum of poles that occur at
\begin{align}
\tan^2 \overline{\xi} = x^{-2}_i - 1\,.
\label{eqn:poles}
\end{align}
In addition, if $m \neq 0$, there is another pole at $\tan^2 \overline{\xi} = 0$.
It is easy to see then that eq~\eqref{eqn:exact_II} may have multiple solutions,
as stated above.
The number of such solutions found in a particular frequency range depends on the permittivities
$\varepsilon_{\perp}(\omega)$ and $\varepsilon_z(\omega)$, which
enter $\tan \overline{\xi}$.
In hBN the hyperbolic response occurs in two separate mid-infrared bands.
The upper band,
$\omega^{\mathrm{TO}}_{\perp} < \omega < \omega^{\mathrm{LO}}_{\perp}$,
is a type II HM~\cite{Guo2012, Poddubny2013}, $\varepsilon_{\perp} < 0$, $\varepsilon_z > 0$.
As frequency $\omega$ changes from $\omega^{\mathrm{TO}}_{\perp}$ to $\omega^{\mathrm{LO}}_{\perp}$,  $\varepsilon_{\perp}$ changes from $-\infty$ to $0$ while $\varepsilon_z$ is positive and approximately constant.
To find the number of the solutions of eq~\eqref{eqn:exact_II}, one just counts the number of the poles crossed by $\tan^2 \overline{\xi}$ as frequency changes.
We can index these solutions by an integer $n$, which is equal to zero if the pole is $\tan^2 \overline{\xi} = 0$ and equal to $i$ if the pole originates from $x_i$, eq~\eqref{eqn:poles}.
One concludes that $n$ runs from $1$ to $N$
for $m = 0$ and from $0$ to $N$ for $m \neq 0$.
At frequencies that belong to the lower band, hBN behaves as a type I HM,~\cite{Guo2012, Poddubny2013} $\varepsilon_{\perp} > 0$, $\varepsilon_z < 0$, and similar analysis yields that $n$ runs from $1$ to $l - |m|- N$.
Therefore, the total number of the solutions in both bands combined is equal to $l$ for $m = 0$ and $l - |m| + 1$ for $m \neq 0$, as stated in the main text.

\subsection{Hamiltonian optics}
\label{sec:HO}

The approximate eigenmodes of our system can also
be found by combining the Hamiltonian optics approach and
the Einstein-Brillouin-Keller (EBK) quantization rules~\cite{Keller1958, Keller1960}.
In this approach the polariton eigenfunctions $\Phi$
are zero modes of the effective bulk Hamiltonian
\begin{equation}
H_B = \varepsilon_{i j} p_i  p_j \,,
\label{eqn:Hamiltonian}
\end{equation}
which describes the region filled by the hyperbolic medium.
The eigenfrequency $\omega$ is contained implicitly in the dielectric tensor
$\varepsilon_{i j}(\omega)$.
It so happens that the boundary condition for $\Phi$ can be written in terms of a single quantity --- the reflection phase shift $\delta$ --- defined below [Eq.~\eqref{eqn:delta}].
This fact leads to the existence of three conserved quantities in the problem,
which implies that the system is integrable.
Two of such conserved quantities are obviously the energy (equal to zero) and the $z$-axis angular momentum $L_z$.
The third conserved quantity $L_{12}$ is
introduced shortly below.

We start with analyzing classical dynamics of the system.
After a canonical transformation to coordinates $(\xi, \theta, \phi)$,
defined by [eq (4) of the main text]
\begin{gather}
\rho = -i b \sqrt{\varepsilon_{\perp}}\,  \sin\xi \sin\theta\,, \quad
z = b \sqrt{\varepsilon_{z}}\, \cos\xi \cos\theta\,,
\label{eqn:inside_coord}\\
b^2 = \varepsilon_z^{-1} a_z^2 - \varepsilon_{\perp}^{-1} a_{\perp}^2\,,
\label{eqn:b}
\end{gather}
where $0 < \xi < {\overline{\xi}}$, ${\overline{\xi}} < \theta < \pi - \overline{\xi}$,
the Hamiltonian $H_B$ becomes [eq~(12) of the main text]
\begin{align}
H_B &= \varepsilon_{i j}\left(\frac{\partial(x,y,z)}{\partial(\xi,\theta,\phi)}\right)
^{\alpha}_{i}
\left(\frac{\partial(x,y,z)}{\partial(\xi,\theta,\phi)}\right)
^{\beta}_{j}  p_{\alpha}  p_{\beta} \notag \\
&= \frac{p_{\xi}^2-p_{\theta}^2}{\sin^2\xi-\sin^2\theta}
- \frac{p_{\phi}^2}{\sin^2\xi \sin^2\theta}\,.
\label{eqn:Hamiltonian_Spheroidal}
\end{align}
The existence of the third conserved quantity $L_{12}$ becomes evident when
one goes through the standard procedure of separation of variables.
In our case, where $H_B = 0$,
the separated expressions for the momenta are
\begin{align}
p_{\xi} &= \pm \sqrt{L_{12}-\frac{L^2_z}{\sin^2\xi}}\,,
\label{eqn:p_xi}
\\
p_{\theta} &= \pm \sqrt{L_{12}-\frac{L^2_z}{\sin^2\theta}}\,,
\label{eqn:p_theta}
\\
p_\phi &= L_z\,.
\label{eqn:p_phi}
\end{align}
The position of the caustic is given by
\begin{align}
	\xi_c = \arcsin \sqrt{\frac{L_z^2}{L_{12}}}\,.
	\label{eqn:xic}
\end{align}
If $0 < \xi_c < \overline{\xi}$,
where $\overline{\xi}$ is given by eq~\eqref{eqn:xi_bar},
the momenta $p_\xi, p_\theta$ are real in the rectangular region $\xi_c \leq \xi \leq \overline{\xi}$,
$\overline{\xi} \leq \theta < \pi - \overline{\xi}$,
see Figure~2b of the main text.
Conversely, if $\xi_c$ exceeds $\overline{\xi}$,
no classically accessible region inside the spheroid exists,
see Figure~2d of the main text.

Note that the velocity of this fictitious motion has components $v_\alpha = \partial H_B / \partial p_\alpha$.
Since $\theta \geq \xi$, the signs of
$v_\theta$ and $p_\theta$ are the same but the signs of $v_\xi$ and $p_\xi$ are opposite.


Equations~\eqref{eqn:p_xi}--\eqref{eqn:p_phi} specify a hypersurface in the six-dimensional phase space that has the topology of a three-dimensional torus \cite{Arnold1989}.
According to the EBK rules, the total phase acquired across any closed loop on this torus should be an integer multiple of $2\pi$.
The phase must include a phase shift of $-\tfrac{\pi}{2}$ upon crossing the caustic and the reflection phase shift(s) at the boundary.
We found (cf.~Sec.~\ref{sec:phase_shift}) that due to the sign structure of the velocity components,
the vertical boundary segment $\xi = \overline{\xi}$ (region $2$ in Fig.~2 of the main text) and the two horizontal ones
$\theta = \overline{\xi}, \pi - \overline{\xi}$
(regions $1$ and $3$ in Fig.~2 of the main text) have opposite phase shifts,
respectively, $\delta$ and $-\delta$.
Therefore, the EBK quantization conditions have the form
\begin{align}
 2 \int\limits_{\xi_c}^{\bar\xi} |p_{\xi}| d\xi -\frac{\pi}{2} + \delta &=  2\pi \nu\,,
\label{eqn:EBK_n}
\\
 2 \int\limits_{\bar\xi}^{\pi-\bar\xi} |p_{\theta}| d\theta - 2 \delta &= 2\pi \lambda\,,
\label{eqn:EBK_l}
\\
 2\pi L_z &= 2\pi \mu\,.
\label{eqn:EBK_m}
\end{align}
Without loss of generality, the quantum numbers $(\mu, \lambda, \nu)$ can be taken to be nonnegative integers.
From eq~\eqref{eqn:EBK_m} we see than $L_z = \mu$.
We now need to express the remaining classical integral of motion $L_{12}$ in terms of the quantum numbers.
Using eqs~\eqref{eqn:p_xi}, \eqref{eqn:p_theta}, \eqref{eqn:EBK_n} and \eqref{eqn:EBK_l}, we obtain
\begin{align}
2 \mu \int\limits_{\xi_c}^{\bar\xi}
 \sqrt{\frac{1}{\sin^2\xi_c} - \frac{1}{\sin^2\xi}}\,\, d\xi -\frac{\pi}{2} + \delta &=  2\pi \nu\,,
\label{eqn:EBK_xi}
\\
2 \mu \int\limits_{\bar\xi}^{\pi-\bar\xi}
 \sqrt{\frac{1}{\sin^2\xi_c} - \frac{1}{\sin^2\theta}}\,\, d\theta  - 2\delta
  &= 2\pi \lambda\,,
\label{eqn:EBK_theta}
\end{align}
which can be rewritten as
\begin{align}
2 \varphi^\mu_\nu({\overline{\xi}}, \xi_c) - \frac{\pi}{2} + \delta = 2 \pi \nu\,,&
\label{eqn:Keller}
\\
4 \varphi^\mu_\nu\left(\frac{\pi}{2}, {\overline{\xi}}\right)
 - 2\delta = 2 \pi \lambda\,.&
\label{eqn:Keller_II}
\end{align}
Here we defined
\begin{align}
 \varphi^\mu_\nu(\xi, \xi_c) &\equiv \mu \int\limits_{\xi_c}^{\xi}
 \sqrt{\frac{1}{\sin^2\xi_c} - \frac{1}{\sin^2\xi}}\,\, d\xi&
 \notag\\
 &= \frac{\mu}{\sin \xi_c} \, A(\xi,\xi_c)
   - \mu B(\xi,\xi_c)\,,
\label{eqn:P_phase}\\
A(\xi,\xi_c) &\equiv \arccos \left(\frac{\cos\xi}{\cos\xi_c} + i 0\right)\,,
\label{eqn:A}
\\
B(\xi,\xi_c) &\equiv \arccos \left( \frac{\cot\xi}{\cot\xi_c} + i 0\right)\,.
\label{eqn:B}
\end{align}
Compared to eq~(24) of the main text, here we add the infinitesimal quantities `$+ i 0$' in the definitions of $A(\xi,\xi_c)$ and $B(\xi,\xi_c)$.
These infinitesimal quantities have no effect at $\xi < \xi_c$ but they will be important in Sec.~\ref{sec:Periodic} where we consider $\xi > \xi_c$ to describe the surface modes.
Combining these equations, we obtain the expression for $\xi_c$:
\begin{equation}
\sin \xi_c = \frac{\mu}{2 \nu + \lambda + \mu + \frac{1}{2}}\,.
\label{eqn:xi_c}
\end{equation}
The final step of the EBK procedure is to account for the boundary conditions,
which entail a certain equation for $\delta$.
This equation can be written as (cf.~Sec.~\ref{sec:phase_shift})
\begin{align}
\tan \frac{\delta}{2} &= i\, \frac{1 - e^{i \delta}}{1 + e^{i \delta}}
\notag\\
  &= \frac{i}{\sqrt{\varepsilon_{\perp}} \sqrt{\varepsilon_z}} \,
              \left(\frac{\dfrac{1}{\sin^2 \xi_c} + \dfrac{1}{\sinh^2 \bar{\eta}}}
                         {\dfrac{1}{\sin^2 \xi_c} - \dfrac{1}{\sin^2 \overline{\xi}}}
              \right)^{\!1 / 2}
     \,,
\label{eqn:delta}
\end{align}
which is the same as eq~(16) of the main text.
Note that $\bar{\eta}$ is determined by the aspect ratio of the spheroid
$a_{\perp} / a_z = \tanh \bar{\eta}$ while $\xi_c$ is fixed by the quantum numbers via eq~\eqref{eqn:xi_c}.
For each given set of these parameters, the system of equations~\eqref{eqn:Keller} and \eqref{eqn:delta} can be solved numerically for $\omega$,
the implicit argument of
$\varepsilon_{\perp}$ and $\varepsilon_z$, to find the desired eigenfrequency.

\subsection{Correspondence between the EBK and the exact eigenmodes}
\label{sec:exact_vs_HO}

We are to compare the following two eigenmode equations.
The first one is from the exact solution, eq~\eqref{eqn:exact}.
The second one is from the EBK method, eqs~\eqref{eqn:Keller},
and \eqref{eqn:delta}.
To do the comparison, we use the asymptotic forms of the associated Legendre functions:
\begin{widetext}
\begin{align}
\mathsf{P}^m_{l}(\cos \theta) &\simeq
\left\{
\begin{array}{cc}
\dfrac{c^m_{l}}
     {\left(\cos^2 \theta_c - \cos^2 \theta\right)^{1/4}}\,
 \cos\left(\varphi^m_l(\theta, \theta_c) -\dfrac{\pi}{4} \right)\,,
  & \theta_c < \theta \leq \dfrac{\pi}{2}\,,\\[1.5em]
\dfrac12 \dfrac{c^m_{l}}
     {\left(\cos^2 \theta - \cos^2 \theta_c\right)^{1/4}}\,
    \exp\Bigl(-\Im \varphi^m_l(\theta, \theta_c) \Bigr)\,,
  & \dfrac{1}{l + \frac{1}{2}} \ll \theta < \theta_c\,,
\end{array}
\right.
\label{eqn:P_asym}\\
c^m_{l} &= (-1)^{m} \left[\frac{2}{\pi}\,
 \frac{1}{l + \tfrac12}\, \frac{(l + m)!}{(l - m)!} \right]^{1/2}\,,
 \quad
\sin \theta_c = \dfrac{m}{l + \frac{1}{2}}\,.
\label{eqn:c}
\end{align}
These expressions are valid for large $l$ and $m$.
They can be derived~\cite{Landauer1951} applying the semiclassical approximation to the Legendre differential equation.
Similarly, for the Legendre function of the second kind one obtains
\begin{align}
Q^m_{l}(\cosh {\eta})  &\simeq \dfrac{\pi}{2}\,
\frac{c^m_{l}}{\left(\cosh^2 {\eta} - \cos^2 \theta_c \right)^{1/4}}
 \exp \left[-\varkappa^m_{l}({\eta}, \theta_c)\right]\,,
\label{eqn:Q_asym}\\
\varkappa^m_{l}({\eta}, \theta_c) &= \frac{m}{\sin \theta_c}\, \cosh^{-1} \left(
   \frac{\cosh {\eta}}{\cos \theta_c} \right)
   - m \sinh^{-1} \left(
      \frac{\coth {\eta}}{\cot \theta_c} \right)\,.
\label{eqn:Q_exp}
\end{align}
The leading contribution to
the logarithmic derivatives of the $\mathsf{P}^m_{l}$ and $Q^m_{l}$ comes from the cosine and the exponential terms, respectively.
Keeping only these terms, we obtain
\begin{align}
\frac{d}{d \xi_c} \,
  \ln \mathsf{P}^{m}_{l}(\cos{\overline{\xi}}) &\simeq
 -\sqrt{\left(l + \frac{1}{2}\right)^2 - \dfrac{m^2}{\sin^2 {\overline{\xi}}}}\,
 \tan\left( \varphi^m_l({\overline{\xi}}, \theta_c) - \frac{\pi}{4}\right)
\,,\\
\frac{d}{d \bar{\eta}}\, \ln Q^{m}_{l}(\cosh \bar{\eta}) &\simeq
-\frac{d}{d \bar{\eta}} \varkappa^m_{l}(\bar{\eta}, \theta_c)
 = -\sqrt{\left(l + \frac{1}{2}\right)^2 + \frac{m^2}{\sinh^2 \bar{\eta}}} \,.
\end{align}
\end{widetext}
Substituting these expressions into eq~\eqref{eqn:exact},
we see that it can be matched with eq~\eqref{eqn:Keller} if we set $\theta_c = \xi_c$, i.e.,
if we make the following correspondence between the two sets of integers:
\begin{equation}
l = 2 \nu + \lambda + \mu,\quad
m = \mu\,.
\end{equation}
The numerical demonstration of this agreement is shown in Fig.~2 of the main article.

\subsection{Hyperbolic surface modes}
\label{sec:Surface}

The momenta inside and outside of the spheroid satisfy the equations
\begin{align}
\frac{p_{\xi}^2-p_{\theta}^2}{\sin^2\xi-\sin^2\theta}
- \frac{p_{\phi}^2}{\sin^2\xi \sin^2\theta}  &=0 && \text{(inside)} \,,
\label{eqn:Two_Hamiltonian_I}
\\
\frac{p_{\eta}^2+p_{\theta}^2}{\cosh^2\eta-\cos^2\theta}
+ \frac{p_{\phi}^2}{\sinh^2\eta \sin^2\theta} &=0 && \text{(outside)} \,,
\label{eqn:Two_Hamiltonian_II}
\\
i\sqrt{\varepsilon_{\perp}} \sqrt{\varepsilon_{z}} p_\xi &= p_\eta && \text{(surface)} \,.
\label{eqn:Two_Hamiltonian_III}
\end{align}
Hyperbolic surface modes (HSM) correspond to imaginary $p_\xi$ and $p_\eta$.
Eliminating these variables from the equations, we obtain
\begin{equation}
H_S = p_{\theta}^2 + \left( \frac{1}{\sin^2\theta} - \frac{1}{\sin^2\xi_c} \right) p_{\phi}^2 = 0  \,,
\label{eqn:Surface_Hamiltonian}
\end{equation}
same as eq~(17) of the main text.
Function $H_S(p_{\theta}, p_{\phi}, \theta)$ can be considered an effective surface Hamiltonian for polaritons.
Here we defined
\begin{equation}
\frac{1}{\sin^2\xi_c}=\frac{1}{\varepsilon_{\perp} \varepsilon_z -1} \left(\varepsilon_{\perp} \varepsilon_z \frac{1}{\sin^2\overline{\xi}} + \frac{1}{\sinh^2\bar{\eta}} \right) \,.
\end{equation}
This formula, valid for the surface waves, replaces eq~\eqref{eqn:xic} for the bulk waves.

\subsection{Uniform approximation}

The lowest-order semiclassical approximation eq~\eqref{eqn:P_asym} diverges near the caustic $\theta = \theta_c$ whereas the actual Legendre function $\mathsf{P}^m_{l}(\cos \xi)$ remains finite.
This is not a serious problem for the bulk modes; however,
for surface waves there is a range of parameters where the boundary $\overline{\xi}$ is close to the caustic $\xi_c$.
This is the cause of the discrepancy between the exact and EBK results seen in Fig.~3 of the main text near the aspect ratio $\mathcal{A} = 1$.
This discrepancy can be greatly reduced by using the uniform approximation~\cite{Thorne1957ael} for the Legendre function:
\begin{align}
\mathsf{P}^m_{l}(\cos \theta) &\simeq \sqrt{\pi} c^m_{l}
\left(
\dfrac{\zeta}{\cos^2 \theta - \cos^2 \theta_c}
\right)^{1/4} \mathop{\mathrm{Ai}}(\zeta)\,,
\label{eqn:P_uniform}\\
\zeta &= e^{i \pi / 3}
\left[-\dfrac32 \varphi^m_l(\theta, \theta_c)\right]^{2 / 3}.
\label{eqn:zeta}
\end{align}
Here $\mathop{\mathrm{Ai}}(\zeta)$ is the Airy function.
If we apply this approximation to the left hand-side of eq~\eqref{eqn:exact},
use $\ln Q^m_{l}(\cosh {\eta})
 \simeq -\varkappa^m_{l}({\eta}, \theta_c)$
on the right-hand side, and keep the leading terms only,
then eq~\eqref{eqn:Two_Hamiltonian_III} gets replaced by
\begin{align}
i\left. \sqrt{\varepsilon_{\perp}} \sqrt{\varepsilon_z} \,
\frac{1}{i}\, \frac{d\zeta}{d{\xi}} \,
\frac{\mathop{\mathrm{Ai}}^{\prime}(\zeta)}{\mathop{\mathrm{Ai}}(\zeta)} \right|_{\xi=\overline{\xi}}
= i\,\left.\frac{\partial}{{\partial}\eta} \varkappa^m_{l}\right|_{\eta=\bar{\eta}}\,.
\label{eqn:uniform_WKB}
\end{align}
Figure 3 of the main text shows an example of applying eq~\eqref{eqn:uniform_WKB} to computing the $(9,2,0)$ surface mode of an hBN spheroid.
It yields an excellent agreement with the exact eigenfrequency of this mode.


\subsection{The phase shift of internal reflections}
\label{sec:phase_shift}

\begin{figure}
\includegraphics[width=3.0in]{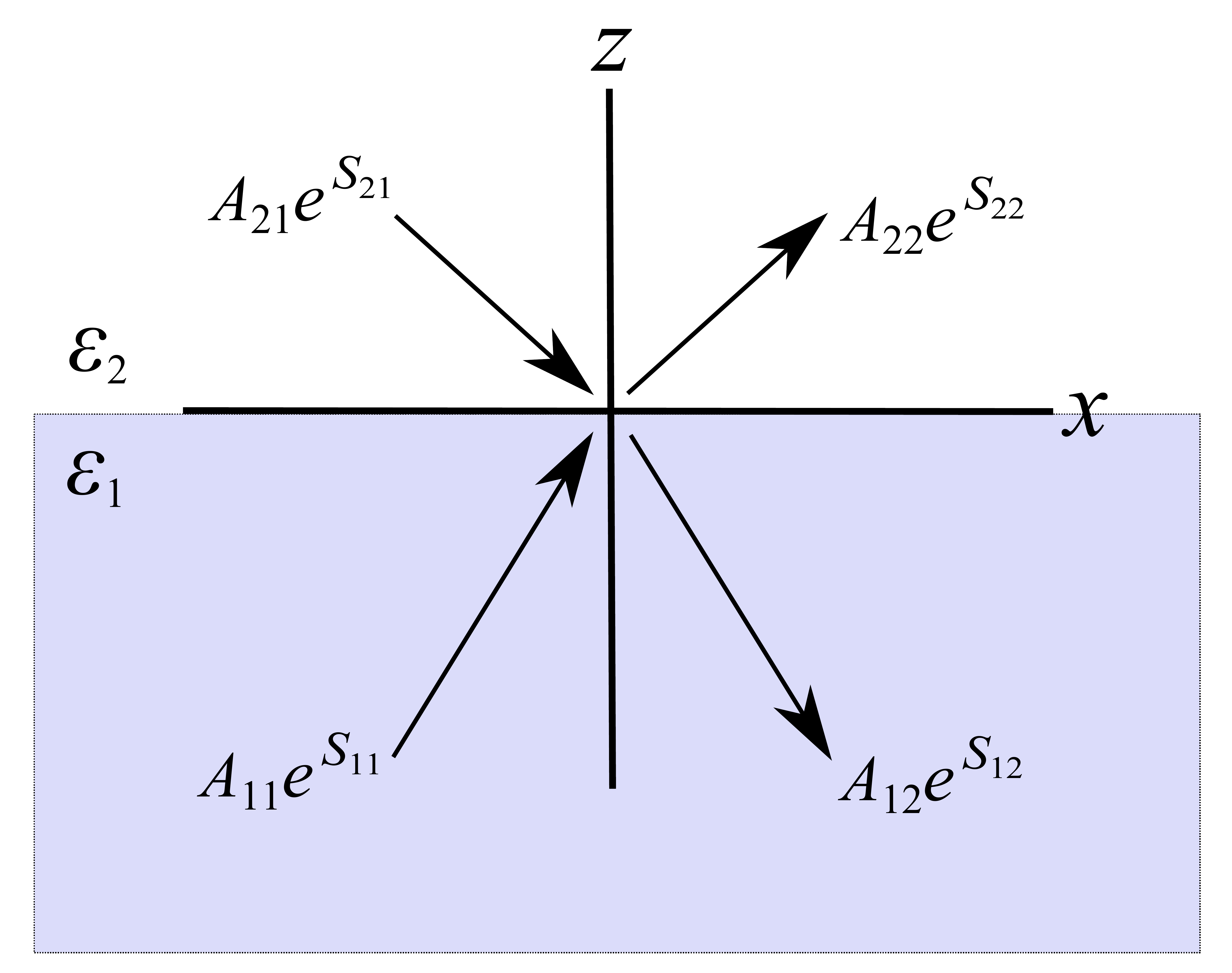}
\caption{Geometry of an auxiliary problem of wave reflection at the boundary of media with permittivity tensors ${\varepsilon}_1$ and ${\varepsilon}_2$.
}
\label{fig:reflection}
\end{figure}

To compute the internal reflection coefficient of polaritons off the spheroid surface, we first consider an auxiliary problem of reflection at the interface of two media, $1$ and $2$, with diagonal dielectric tensors $\hat{\varepsilon}_1$ and $\hat{\varepsilon}_2$, respectively, see Figure~\ref{fig:reflection}.
To solve this latter problem we write the scalar potentials of incident and reflected waves as follows:
\begin{align}
\Phi_1&=A_{11} e^{S_{11}} + A_{12} e^{S_{12}} \,, \\
\Phi_2&=A_{21} e^{S_{21}} + A_{22} e^{S_{22}} \,.
\end{align}
The first index in the phases $S_{j k}$ labels the medium, and the second distinguishes incident $k = 1$ and reflected $k = 2$ waves.
The gradients of the phases $S_{j k}$ are the wave momenta $p^\alpha_{j k} = -i \partial_\alpha S_{j k}$.
We assume that only $\alpha = z$ and $x$ components are nonzero.
The boundary conditions are:
\begin{equation}
\Phi_1 = \Phi_2 \,, \quad
\varepsilon_{1z} \partial_z \Phi_1 = \varepsilon_{2z} \partial_z \Phi_2 \,.
\label{boundary_condition}
\end{equation}
To satisfy them, all $p^x_{j k}$'s must be equal.
In addition, we must have
\begin{align}
A_{11} + A_{12} &= A_{21} + A_{22} \,, \notag\\
A_{11} \partial_z S_{11} + A_{12} \partial_z S_{12} &=
\frac{\varepsilon_{2z}}{\varepsilon_{1z}}( A_{21} \partial_z S_{21} + A_{22} \partial_z S_{22})\,.
\notag
\end{align}
Taking advantage of the fact that $\partial_z S_{11}=-\partial_z S_{12}$ and $\partial_z S_{21}=-\partial_z S_{22}$, we get
\begin{align}
\begin{pmatrix}
A_{11} \\
A_{12}
\end{pmatrix}
=
\frac{1}{2}
\begin{pmatrix}
1+t & 1-t\\
1-t & 1+t
\end{pmatrix}
\begin{pmatrix}
A_{21} \\
A_{22}
\end{pmatrix}\,,
\end{align}
where
\begin{equation}
t = \frac{\varepsilon_{2z}}{\varepsilon_{1z}}\,
    \frac{\partial_z S_{21}}{\partial_z S_{11}} \,.
\end{equation}
Setting $A_{21}=0$, which means there is only outgoing wave in medium $2$,
we get the reflection coefficient
\begin{equation}
e^{i\delta} = \frac{1+t}{1-t}
\label{phase_shift_supplemental}\,.
\end{equation}
If the wave in medium $2$ is evanescent, i.e., if momenta $p^z_{2 k}$ are imaginary,
then the reflection phase shift $\delta$ is real.

Next, we turn to our original problem of internal polariton reflection at the surface of a suspended nanogranule.
The problem can be reduced to the one solved above using the special
choice of coordinates: coordinates $(\xi, \theta, \phi)$ inside the granule [eq~\eqref{eqn:inside_coord}] and the spheroidal coordinates $(\eta, \theta, \phi)$ in vacuum outside [eq~(2) of the main text].
Equation~\eqref{boundary_condition} becomes:
\begin{equation}
\Phi_1 = \Phi_2, \quad
i \sqrt{\varepsilon_{\perp}}\sqrt{\varepsilon_z}\,
 \partial_\xi \Phi_1 = \partial_\eta \Phi_2 \,.
\label{eqn:bc}
\end{equation}
Therefore, eq.~\eqref{phase_shift_supplemental} holds after the following trivial change is made:
\begin{equation}
t = \frac{1}{i\sqrt{\varepsilon_{\perp}}\sqrt{\varepsilon_z}}\,
    \frac{\partial_\eta S_{21}}{\partial_\xi S_{11}}
 = \frac{1}{i\sqrt{\varepsilon_{\perp}}\sqrt{\varepsilon_z}}\,
   \frac{p_\eta}{p_\xi} \,.
\end{equation}
Taking advantage of eq~\eqref{eqn:p_xi} for $p_\xi$ and the similar expression for
momentum $p_\eta$ outside the spheroid,
\begin{equation}
p_\eta = i\, \sqrt{L_{12} + \frac{L^2_z}{\sinh^2 \eta}}\,,
\end{equation}
we recover eq~\eqref{eqn:delta}.

\section{Periodic Orbits}
\label{sec:Periodic}

\begin{figure*}
\begin{center}
\begin{minipage}{2.2in}
\begin{flushright}
\noindent
{\includegraphics[width=0.4in]{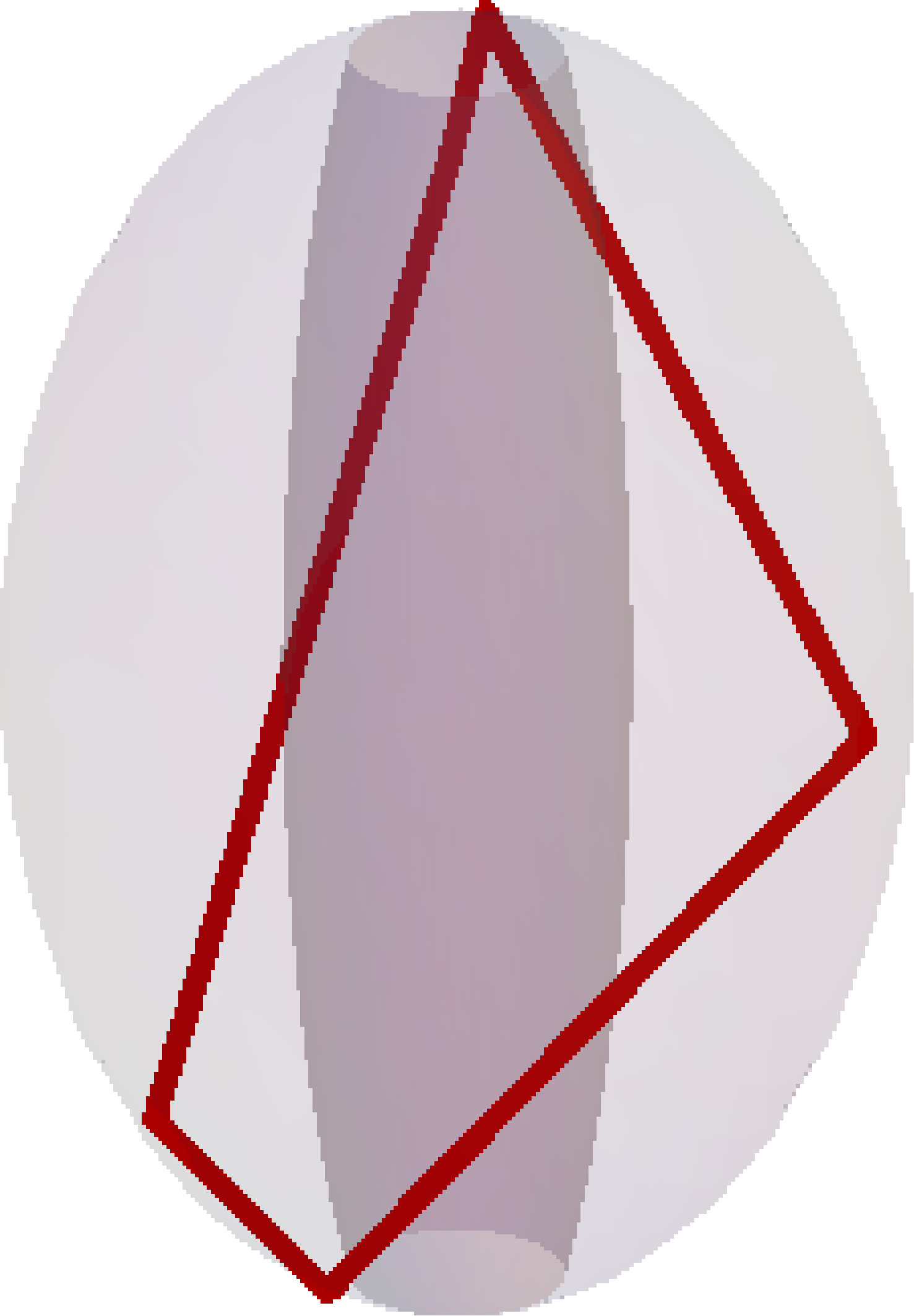}}\hspace{0.05in}
{\includegraphics[width=0.4in]{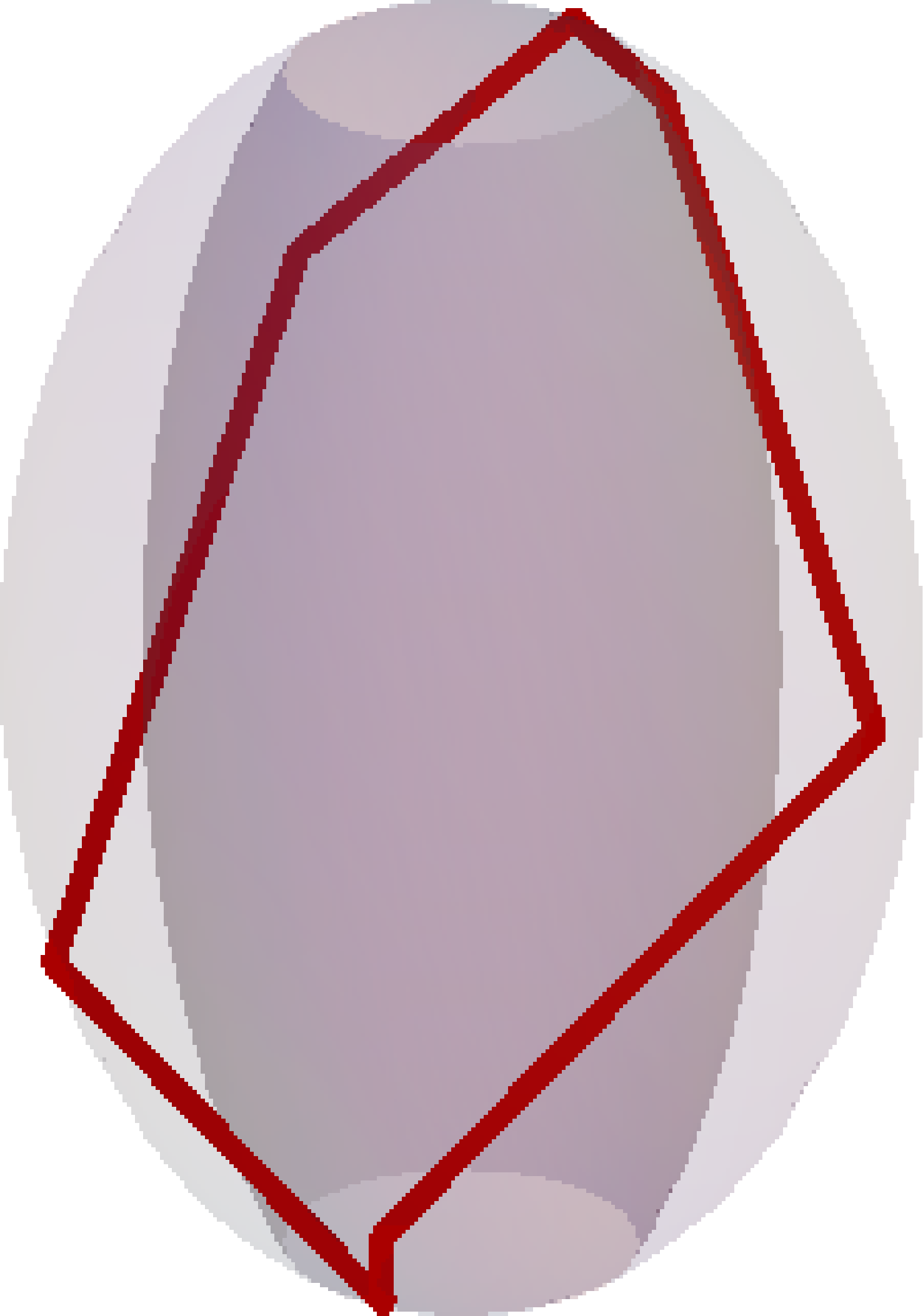}}\hspace{0.05in}
{\includegraphics[width=0.4in]{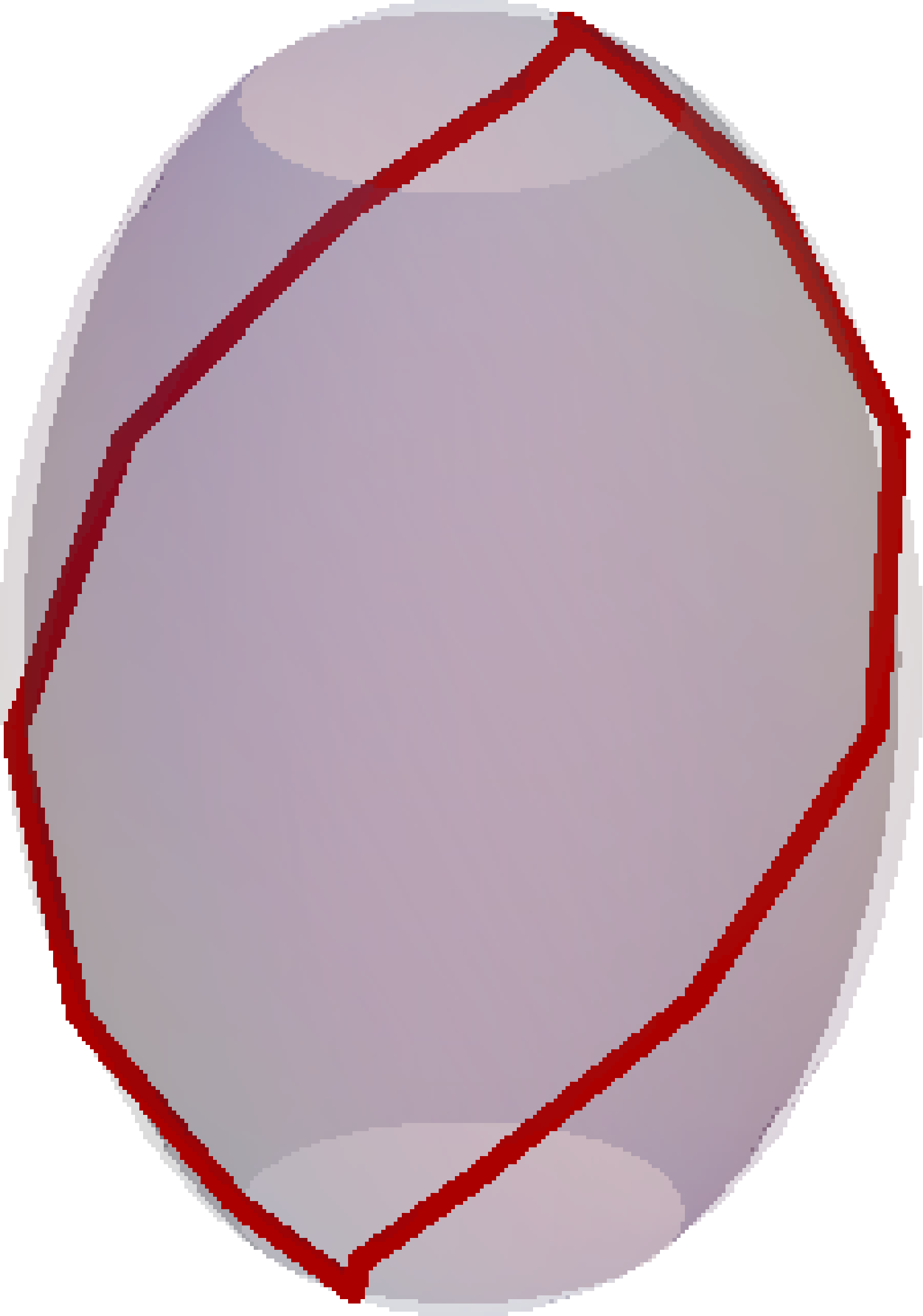}}\hspace{0.05in}
{\includegraphics[width=0.4in]{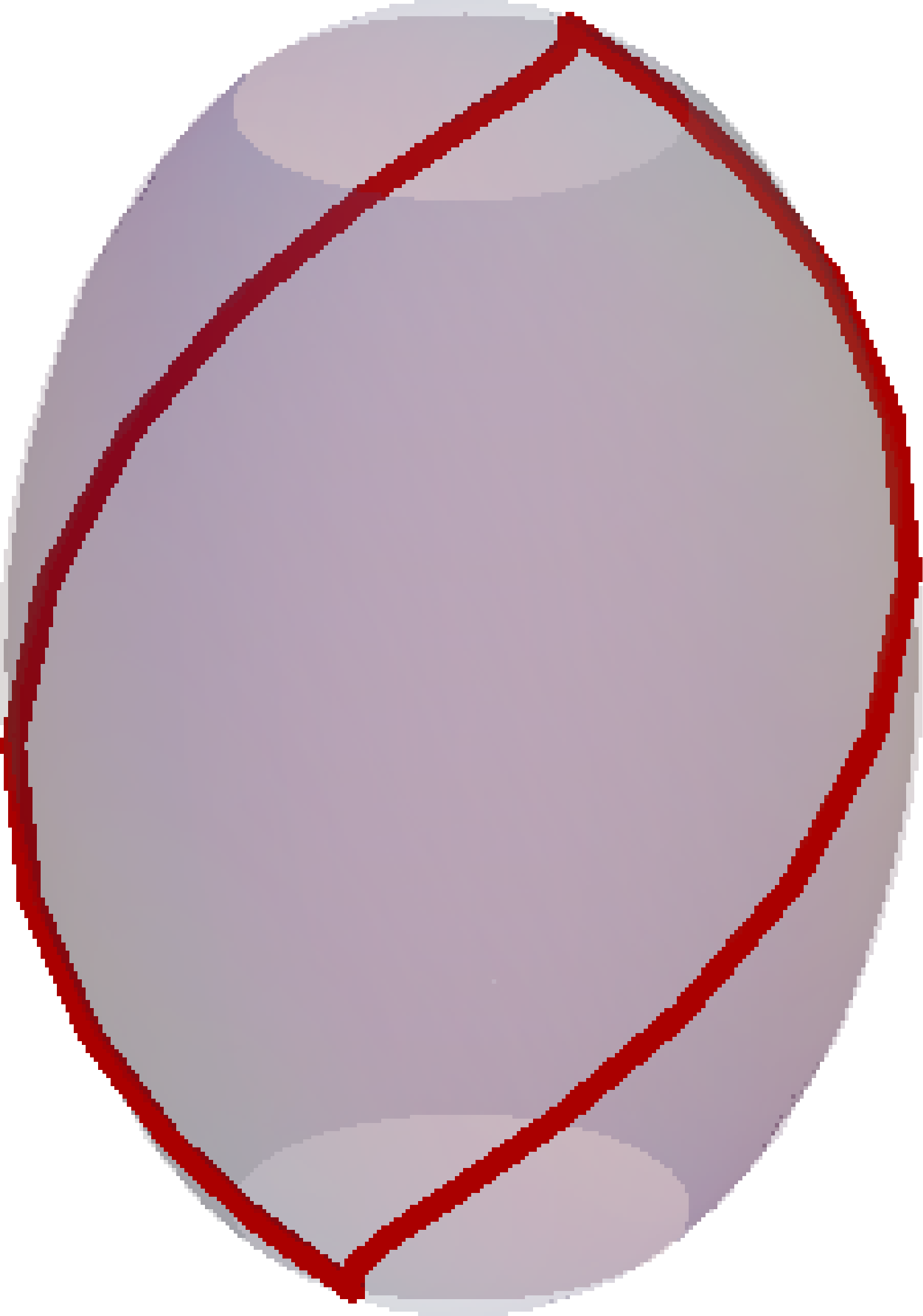}}
\newline
\noindent
{\includegraphics[width=0.4in]{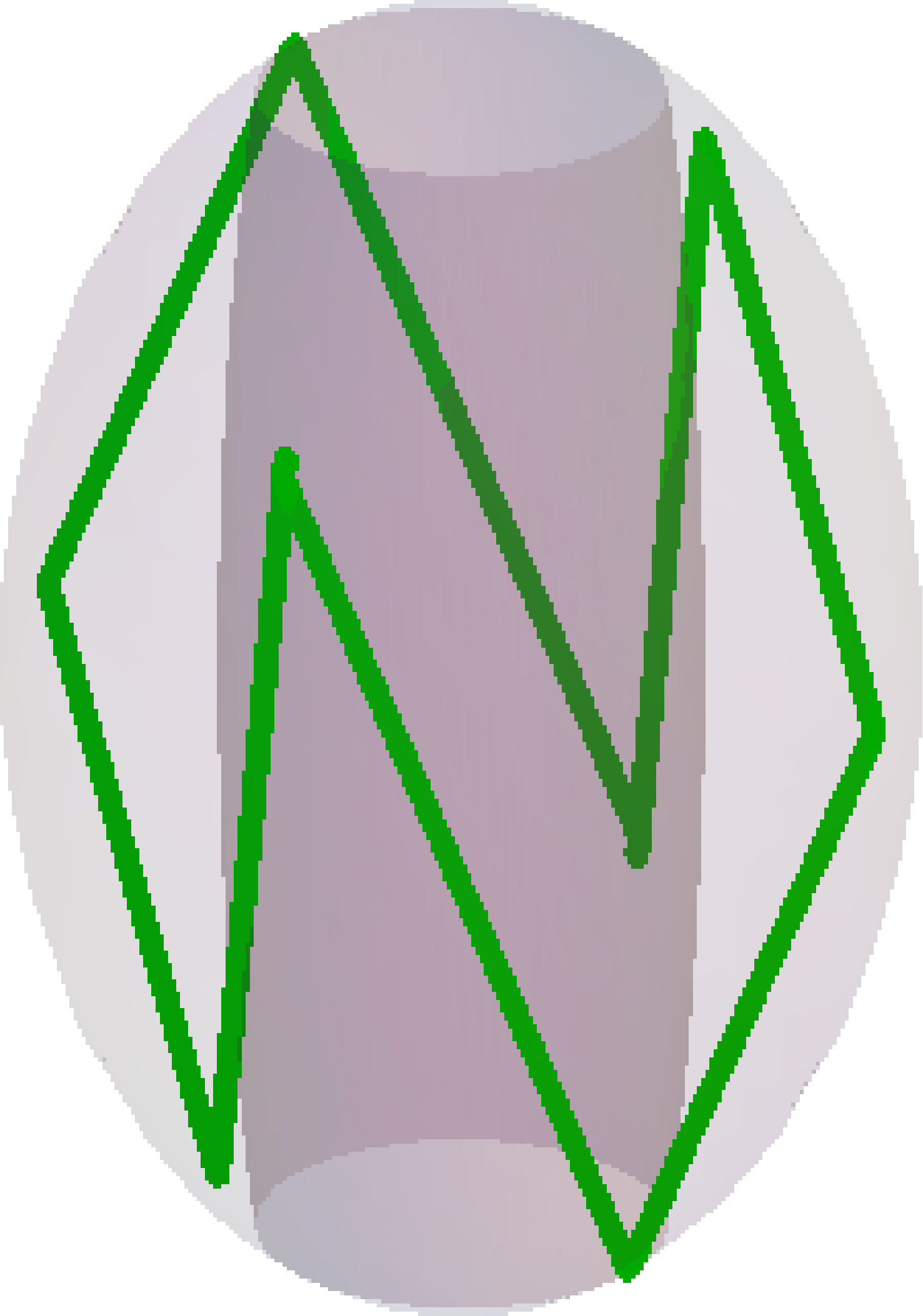}}\hspace{0.05in}
{\includegraphics[width=0.4in]{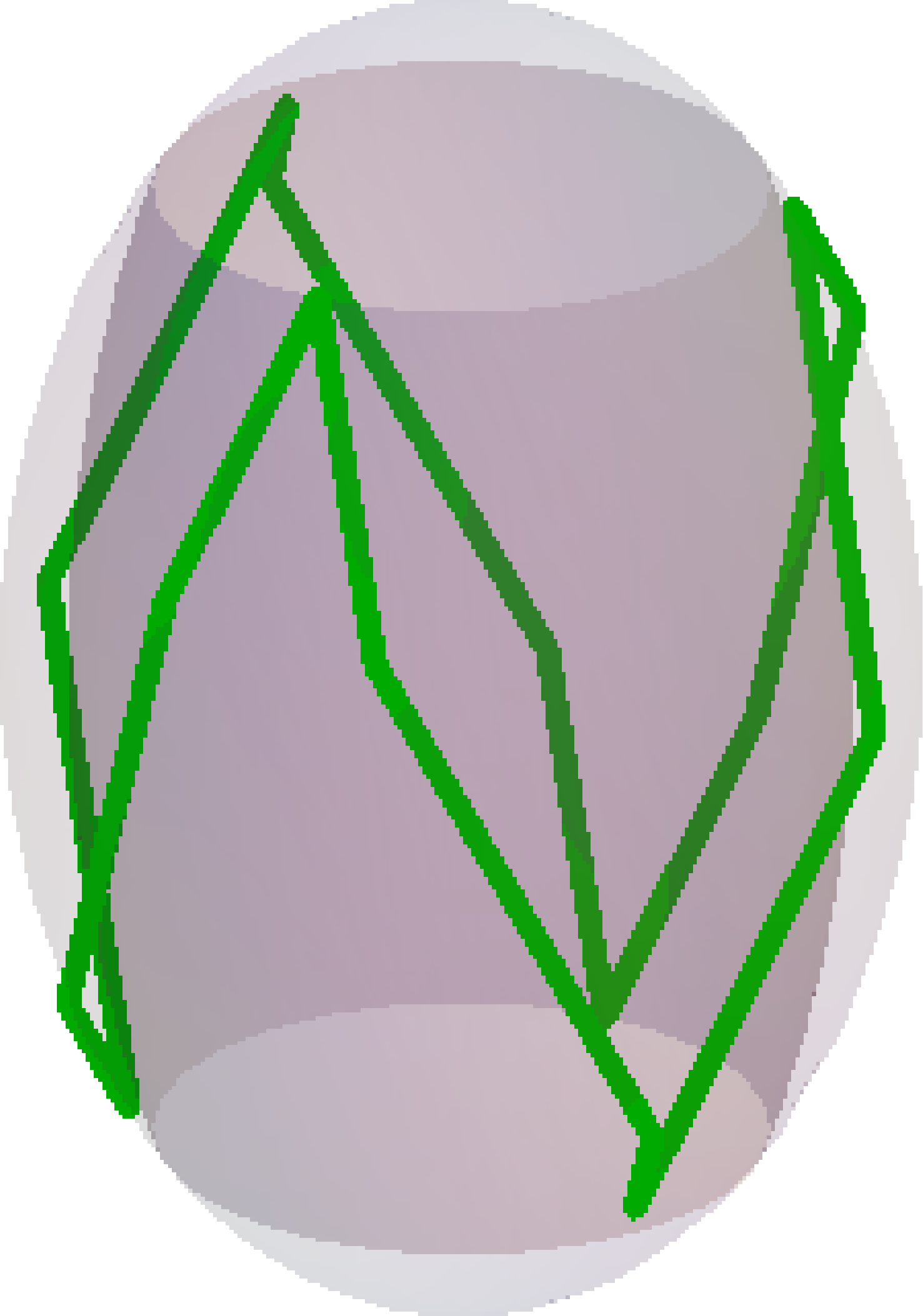}}\hspace{0.05in}
{\includegraphics[width=0.4in]{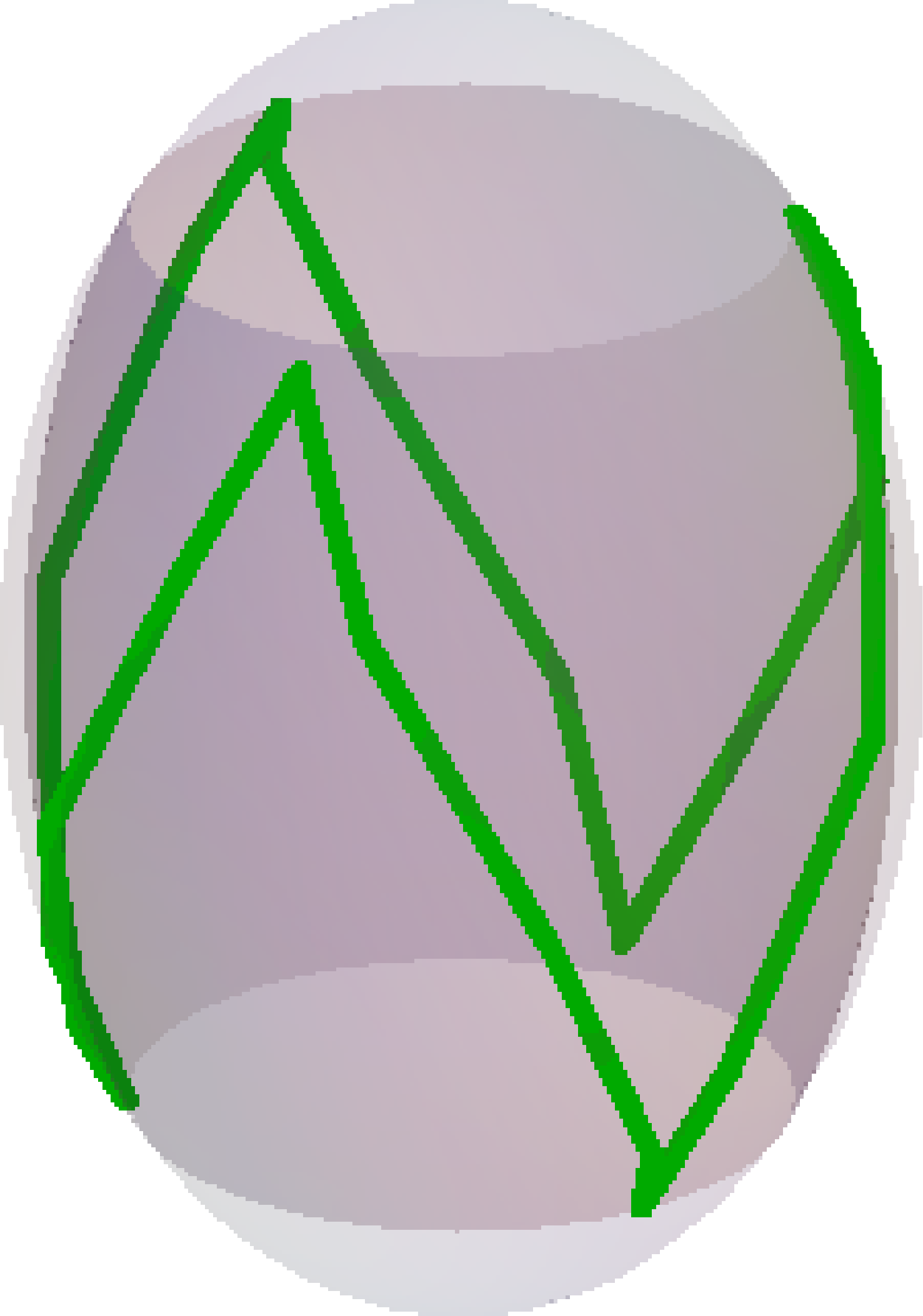}}\hspace{0.05in}
{\includegraphics[width=0.4in]{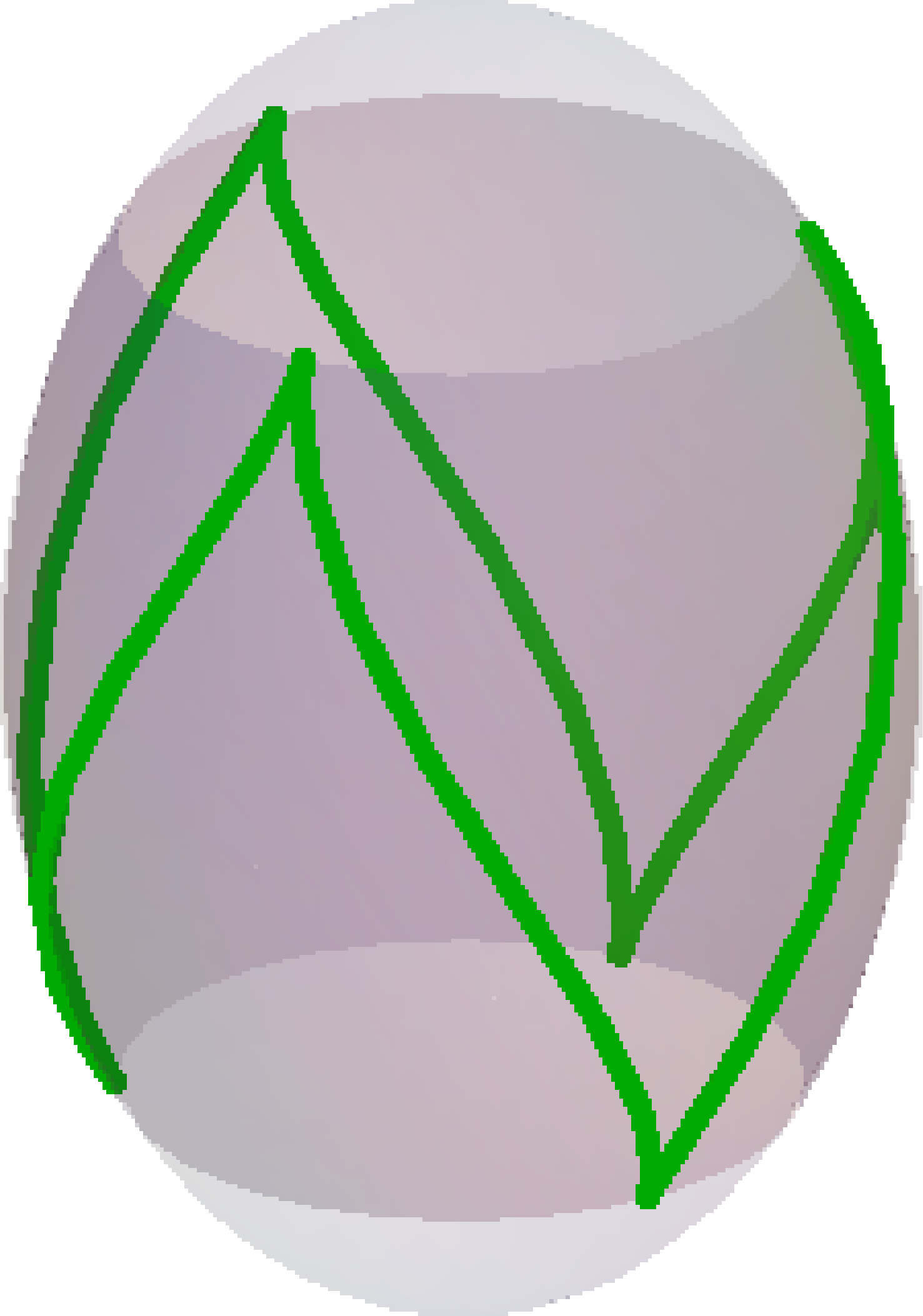}}
\newline
\noindent
{\includegraphics[width=0.4in]{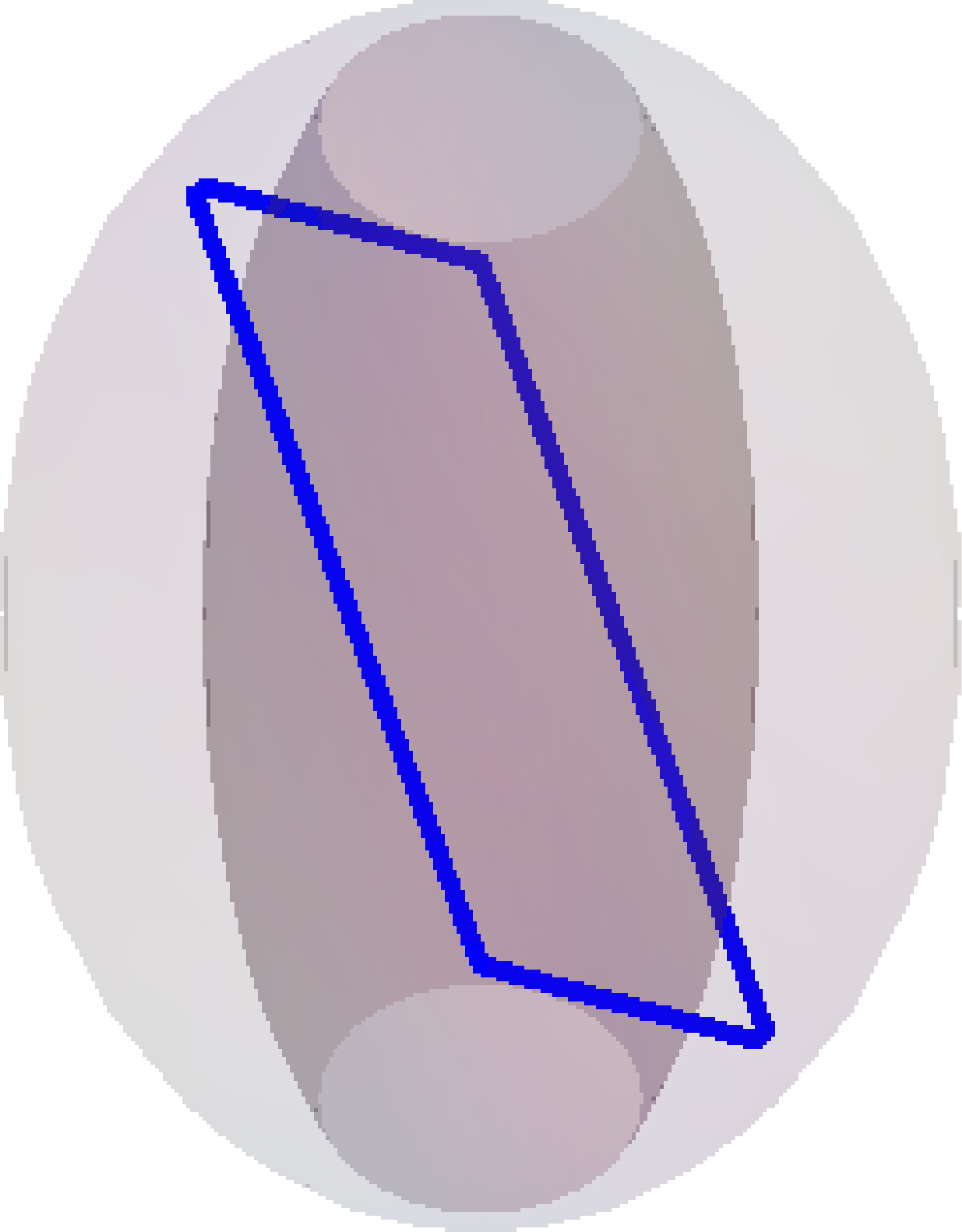}}\hspace{0.05in}
{\includegraphics[width=0.4in]{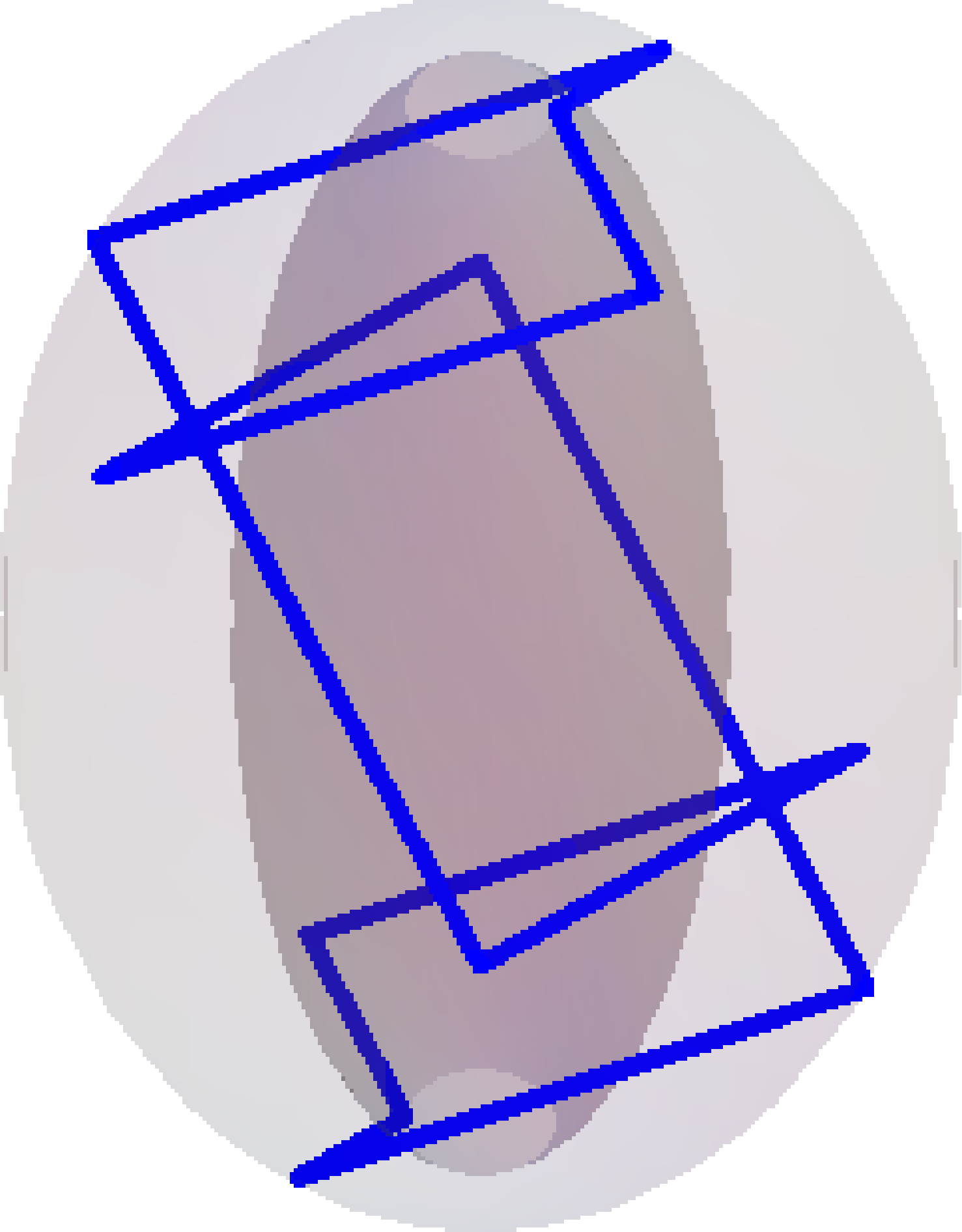}}\hspace{0.05in}
{\includegraphics[width=0.4in]{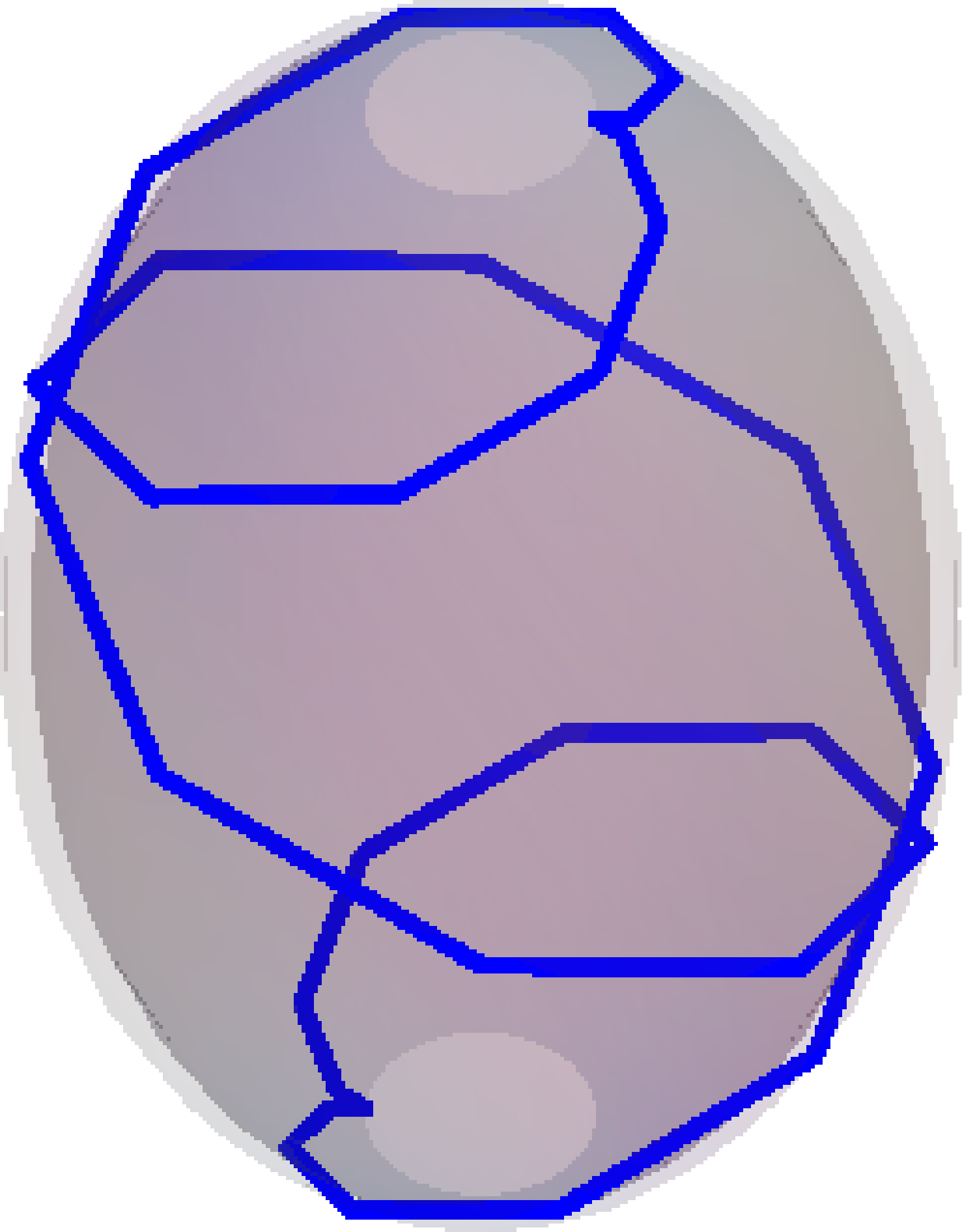}}\hspace{0.05in}
{\includegraphics[width=0.4in]{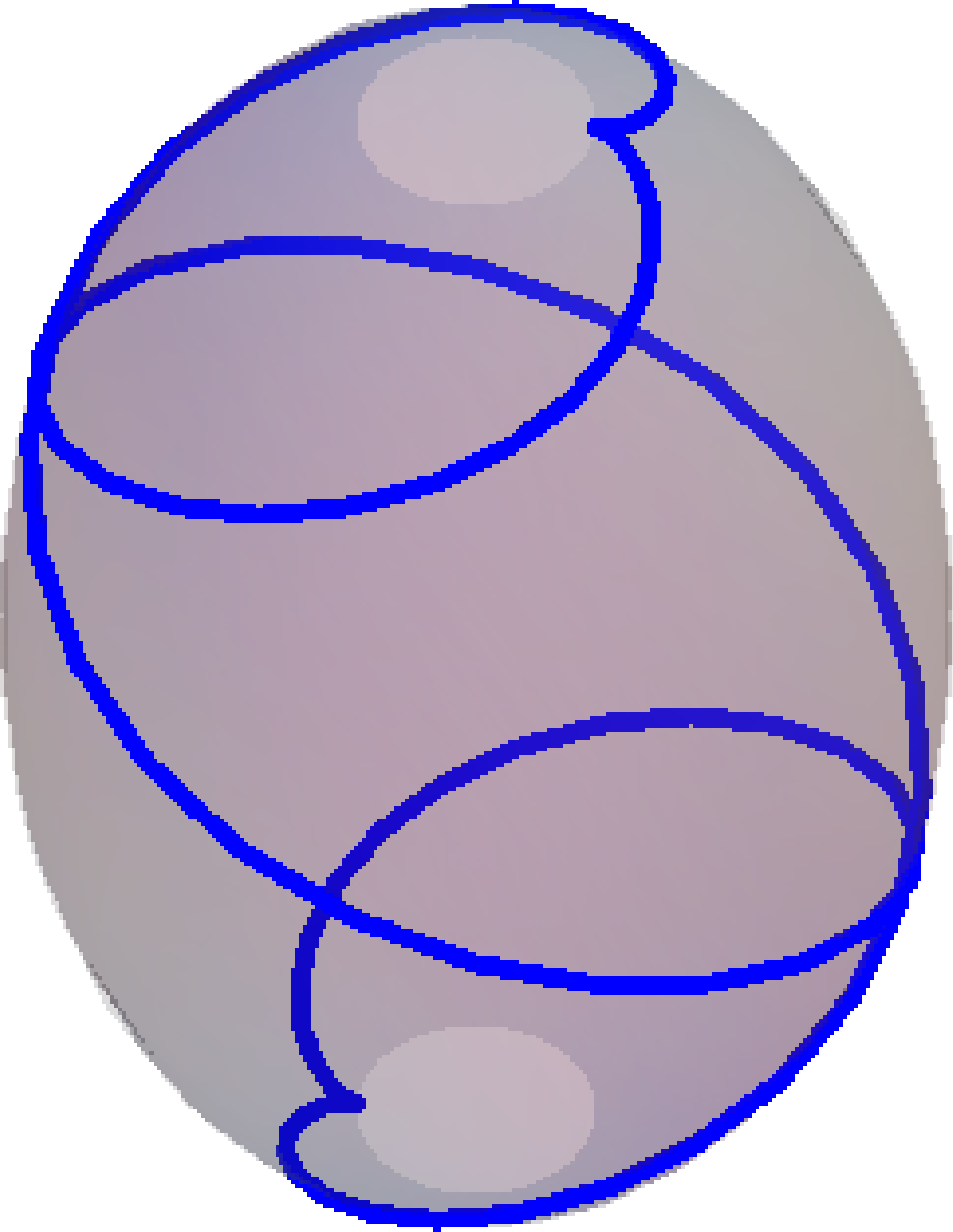}}
\newline
\end{flushright}
\end{minipage}
\begin{minipage}{4.2in}
\raisebox{-1.2\height}{\includegraphics[width=2.8in]{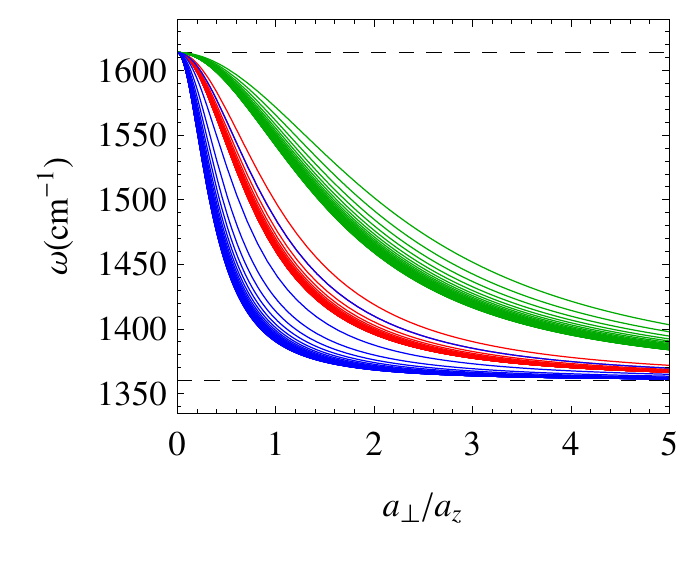}}
\raisebox{-1.2\height}{\includegraphics[width=1.3in]{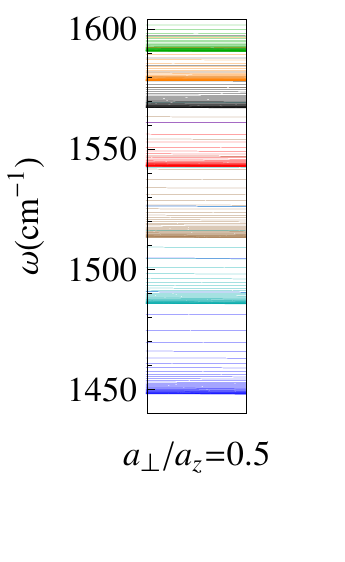}}
\end{minipage}
\end{center}
\caption{(Color online)
Left: periodic orbits for different values of $(\tau_\xi, \tau_\theta, \tau_\phi)$.
The red, green, and blue correspond to, respectively, $|\tau_\theta / \tau_\phi| = 1$, $1 / 3$, and $3$.
Parameter $|\tau_\xi|$ decreases from left to right.
Smaller $\tau_\xi$'s yield smoother orbits.
The value of $|\tau_\xi / \tau_\phi|$ can be inferred from the number of radial bounces that occur during one azimuthal ($\phi$) cycle.
Center: the dispersions (electromagnetic field frequencies) of the periodic orbits as a function of the aspect ratio $\mathcal{A} = a_\perp / a_z$ of the granule. The same color code as in the left panel is used. As $|\tau_\xi|$ decreases, the distance between adjacent curves of the same color decreases. The $\tau_\xi \to 0$ limit corresponds to whispering gallery trajectories grazing along the surface of the granule.
The first $40$ inverse integer values of $|\tau_\xi|$ are included.
Right: the periodic orbits dispersions for $\mathcal{A} = 0.5$. The orbits include $|\tau_\theta / \tau_\phi| = i / j$, where $\{i, j\} \in \{1, 2, 3\}$.
The values of $|\tau_\xi|$ are the same as in the central panel.
}
\label{fig:bunching}
\end{figure*}

To study classical periodic orbits of the polaritons,
it is convenient to perform a canonical transformation to action-angle
variables.~\cite{Arnold1989}
The new momenta are the actions of the three
independent loops on the three-dimensional torus specified by the constants of motion, cf.~eqs~\eqref{eqn:p_xi}--\eqref{eqn:p_phi}:
\begin{align}
&J_\xi = \oint p_{\xi} d\xi = 2 \int\limits^{\overline{\xi}}_{\xi_c} \sqrt{L_{12}-\frac{L^2_z}{\sin^2\xi}}\, d\xi \,, \\
&J_\theta = \oint p_{\theta} d\theta = 2 \int\limits^{\pi-\overline{\xi}}_{\overline{\xi}} \sqrt{L_{12}-\frac{L^2_z}{\sin^2\theta}}\, d\theta \,, \\
&J_\phi = \oint p_{\phi} d\phi=2\pi L_z \,.
\end{align}
The motion of each action-angle coordinate pair is periodic
with the period
\begin{align}
\tau_i = \left(\frac{\partial H_B}{\partial J_i}\right)^{-1},
\quad i = \xi, \theta, \phi\,.
\end{align}
As explained in the main text, polariton wavepackets follow the same trajectories in the real space and thus also in the space of angle variables as the fictitious particle with Hamiltonian $H_B$. However, the rate of change of the angle variables for polariton wavepackets is different from $\partial H / \partial J_i$ by a certain overall factor.
Therefore, the periods $\tau_i$ themselves do not have a direct physical meaning but their ratios do.
The phase-space trajectory or ``orbit'' is closed if these ratios are rational numbers.

Each $\tau_i$ can be represented by a certain Jacobian.
For example, $\tau_\xi$ is given by
\begin{align}
\tau_\xi^{-1} &= \frac{\partial H_B}{\partial J_\xi} = \frac{\partial(H_B,J_\theta,J_\phi)}{\partial(J_\xi,J_\theta,J_\phi)} \notag\\
&= \frac{\partial(H_B,J_\theta,J_\phi)}{\partial(H_B,L_{12},L_z)}\, \frac{\partial(H_B,L_{12},L_z)}{\partial(J_\xi,J_\theta,J_\phi)} \notag\\
&= \frac{\partial(H_B,J_\theta,J_\phi)}{\partial(H_B,L_{12},L_z)}  \left[
\frac{\partial(J_\xi,J_\theta,J_\phi)}{\partial(H_B,L_{12},L_z)}
\right]^{-1} \notag\\
&= \left(\frac{\partial J_\theta}{\partial L_{12}}\right)_{L_z, H_B}
\left[\frac{\partial(J_\xi,J_\theta)}{\partial(H_B,L_{12})}\right]^{-1} \,.
\end{align}
For the other two periods, $\tau_\theta$ and $\tau_\phi$, we obtain
\begin{align}
\tau_\theta^{-1} &=
\left(\frac{\partial J_\xi}{\partial L_{12}}\right)_{L_z, H_B}
\left[\frac{\partial(J_\theta,J_\xi)}{\partial(H_B,L_{12})}\right]^{-1}, \\
\tau_\phi^{-1} &= \frac{1}{2\pi}\, 
\frac{\partial (J_\xi,J_\theta)}{\partial(L_{12},L_z)}
\left[\frac{\partial(J_\xi,J_\theta)}{\partial(H_B,L_{12})}\right]^{-1} \,.
\end{align}
Their ratios can be reduced to the following form:
\begin{align}
\tau_\xi^{-1}: \tau_\theta^{-1}: \tau_\phi^{-1}
 = \frac{\partial J_\theta}{\partial L_{12}}
  : -\frac{\partial J_\xi}{\partial L_{12}}
  : \frac{1}{2\pi} \frac{\partial (J_\xi,J_\theta)}{\partial(L_{12},L_z)} \,,
\end{align}
where the first two derivatives are to be taken at fixed $L_z$ and fixed $H_B = 0$. After some algebra, we obtain the explicit formulas
\begin{align}
\tau_\xi^{-1}: \tau_\theta^{-1}: \tau_\phi^{-1} = \pi - 2A : -A : (B - A) \,\mathrm{sgn}\, L_z\,,
\label{eq:frequency_ratio}
\end{align}
where $A = A(\overline{\xi},\xi_c)$ and $B = B(\overline{\xi},\xi_c)$ are defined by
eqs.~\eqref{eqn:A} and \eqref{eqn:B}.
To get a particular periodic orbit, we follow these steps.
First, we choose the period ratios to be desired
rational numbers.
Next, we determine $A$ and $B$ consistent with this choice.
Next, we solve for the constants of motion $L_z$ and $L_{12}$
from eqs.~\eqref{eqn:A} and \eqref{eqn:B}.
Finally, the orbit is generated and plotted using eqs.~\eqref{eqn:p_xi}--\eqref{eqn:p_phi}.
In general, the orbits can have very complicated shapes,
as illustrated in Figure~\ref{fig:bunching} (left).
Roughly speaking, the ratio $|\tau_\phi / \tau_\theta|$ determines the topology or the winding number of the orbit whereas $|\tau_\xi|$ determines the typical radial distance of the orbit from the center of the spheroid.
As $|\tau_\xi|$ decreases, the orbit is pushed closer towards the surface of the spheroid.
In the limit $\tau_\xi \to 0$, the
orbit becomes a smooth trajectory grazing along this surface.
This kind of trajectories are similar to the whispering gallery modes well known in ray optics and acoustics.
Therefore, they can be considered a generalization of the whispering gallery modes to the present case of the indefinite Hamiltonian $H_B$.
For positive-definite Hamiltonians it has been rigorously proven~\cite{Lazutkin1993KAM, Amiran1997isp} that the motion along trajectories sufficiently close to the surface of a smooth billiard is regular.
Therefore, such whispering gallery modes are subject to the EBK quantization rules.~\cite{Keller1960}
We expect that the same property holds for indefinite Hamiltonians as well.

%

In fact, a precise relation between classical periodic orbits and quantization should exist. 
According to the trace formulas given by Gutzwiller~\cite{Gutzwiller.1990} for chaotic Hamiltonian systems and by Berry and Tabor~\cite{Berry.1977} for integrable ones, the density of 
states (DOS) of the quantized eigenmodes can be represented by a sum over the periodic orbits.
However, in the present case of the indefinite Hamiltoninan, although the ray dynamics in a spheroidal particle is of course integrable, the density of states (DOS) is divergent without a momentum cutoff.
Therefore, if one carries out the summation in the Berry-Tabor formula~\cite{Gutzwiller.1990, Berry.1977},
one should get infinity not only at some discrete frequencies that are equal to resonance frequencies but in fact at all frequencies in the Restshrahlen band.
How to treat these divergencies is an intriguing problem for future work.

\begin{figure}
\begin{center}
\includegraphics[width=3.2in]{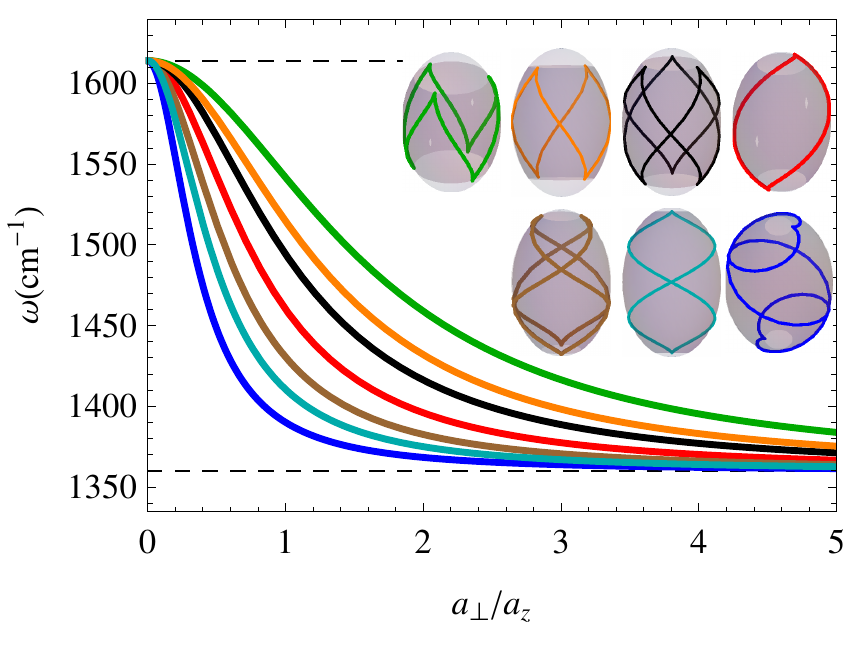}
\end{center}
\caption{(Color online).
Frequencies of the whispering gallery periodic orbits as functions of the aspect 
ratio $\mathcal{A}$.
The orbits have period ratios $|\tau_\theta / \tau_\phi| = i/j$ with $\{i,j\}=\{1,2,3\}$ (same as in Figure \ref{fig:bunching}, right).
The inset illustrates the shape of the orbits in the real space.
}
\label{fig:limits}
\end{figure}

The role of short periodic orbits in our system is also unconventional.
A fair approximation to the exact DOS of a billiard system
with the usual quadratic Hamiltonians $H = p^2 / 2 m$,
can be obtained including only the contributions of the shortest orbits.~\cite{Berry.1977}
However, in our case no obvious features of the eigenmode spectra near the electromagnetic frequencies $\omega$ corresponding to short periodic orbits can be identified.
We speculate that the geometric length of the orbit may not be a relevant quantity for systems with indefinite Hamiltonians such as $H_B$.
Note that the frequencies $\omega$ of families of orbits having the same 
$|\tau_\phi / \tau_\theta|$ and decreasing $|\tau_\xi|$ tend to cluster together.
The corresponding frequencies converge to certain value that is a function of the aspect ratio $\mathcal{A} = a_{\perp} / a_{z}$, see the central panel in Figure \ref{fig:bunching}.
Such $\omega$ are plotted in Figure~\ref{fig:limits}.
Naively, a high density of periodic orbits near these whispering galley frequencies may lead to enhancement of the properly regularized DOS.
This problem remains to be understood.

As discussed in Sec.~\ref{sec:Surface}, the momentum $p_\xi$ can be imaginary for large enough angular momenta $L_z$ or at small enough aspect ratios $\mathcal{A}$.
Such waves are not the whispering gallery waves.
Instead, they are HSM described by the surface Hamiltonian $H_S$.
The analysis of the periodic orbits of the HSM is simpler because
there is only one period ratio,
\begin{align}
{\tau_\phi^{-1}}: {\tau_\theta^{-1}}
= 1 - \frac{1}{\sin{\xi_c}}\,.
\label{eqn:surface_per_orbit}
\end{align}
The HSM orbit is closed if $\tau_\phi^{-1}: {\tau_\theta^{-1}} = n_1: n_2$ where $n_1$ and $n_2$ are integers.
Dispersion of several such orbits as a function of $\mathcal{A}$ are shown in Figure~4 of the main text.
Comparing eq~\eqref{eqn:surface_per_orbit} with the EBK condition
\begin{equation}
1 - \frac{1}{\sin{\xi_c}} = -\frac{2\lambda + 1}{2 \mu}\,,
\label{eqn:xi_c}
\end{equation}
which follows from eq~(19) of the main text, we see that roughly a quarter of all possible HSM periodic orbits (those with odd $n_1$ and even $n_2$) are simultaneously EBK eigenmodes.

\section{Response to a dipole}
\label{sec:Dipole}

\subsection{Quasi-static approximation}

\begin{figure}
	\includegraphics[width=3.2in]{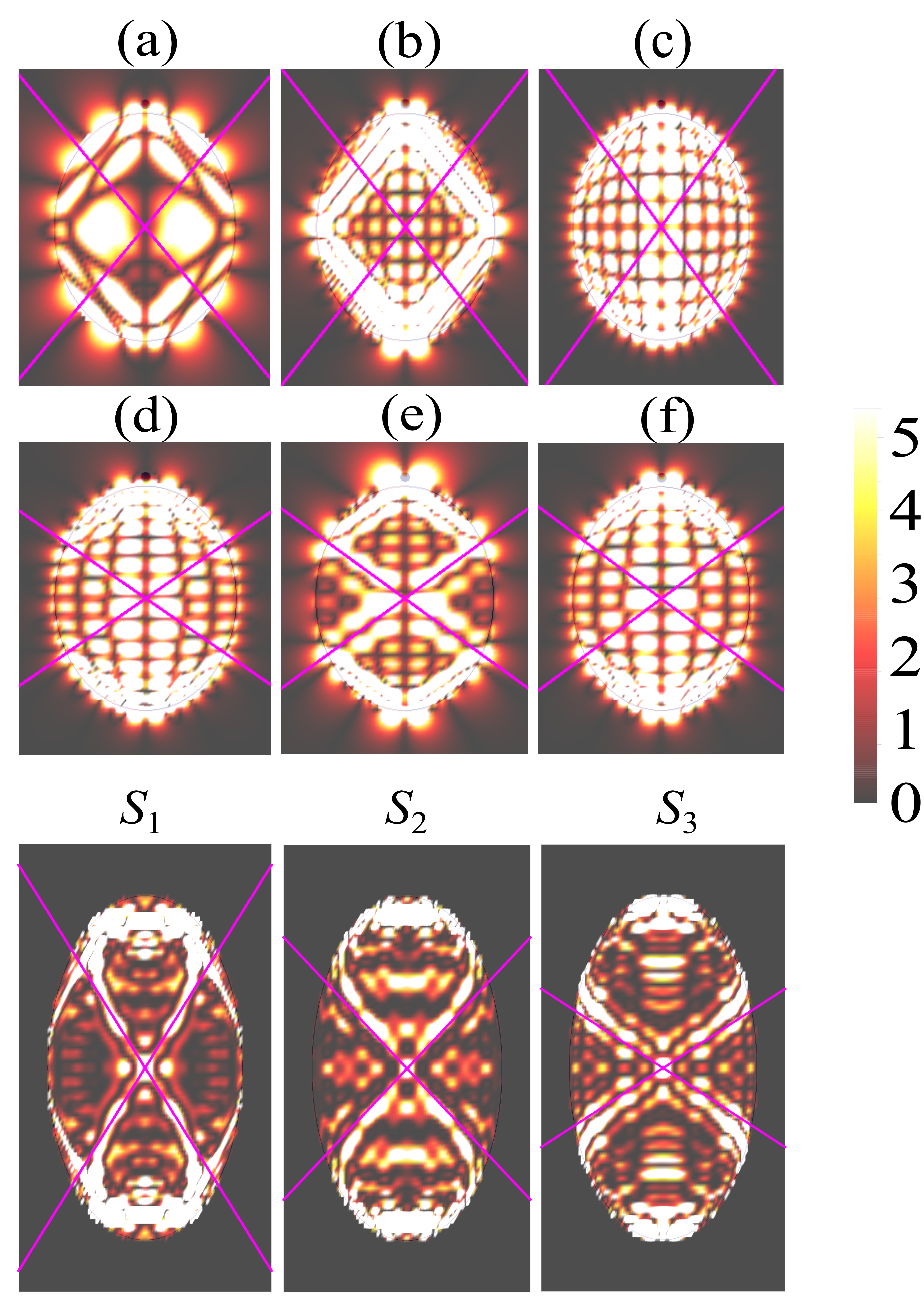}
	\caption{(a)--(f): False color plot of $|E_x|$ in a meridional cross section of a hBN spheroid due to a dipole source located just above the north pole and pointed north. Damping loss is neglected. Outside the spheroid, the dipole's own field is subtracted away, for clarity.
		The frequencies in (a)--(c) are $1551$, $1555$, and $1561\,\text{cm}^{-1}$. In (d)--(f), they are $1490$, $1494$, and $1497 \,\text{cm}^{-1}$. The middle numbers in these sets match the frequencies of $B_1$ and $B_2$ in Figure~4 of the main text for the chosen aspect ratio $\mathcal{A} = \tanh 1 \approx 0.761$.
		The tilted magenta lines run parallel to the polariton group velocity.
		$S_1$--$S_3$: False color plot of $E^2_z$ at the surface of the spheroid projected onto the meridional plane. The dipole is just above the surface at the center of
		each plot. The frequencies in $S_1$--$S_3$ are $1557$, $1535$, $1488\,\text{cm}^{-1}$, same as in Figure~4 of the main text for the chosen aspect ratio $\mathcal{A} = \tanh 0.5 \approx 0.462 $. The tilted magenta lines run parallel to the HSM group velocity at the center of the image.}
	\label{fig:field_bulk}
\end{figure}

In this section we outline the steps needed to calculate the field created by the nanogranule in response to a nearby oscillating electric dipole.
We assume that the dipole is located at a point $\textbf{R}$ in the $x$--$z$ plane.
Let the spheroidal coordinates of $\textbf{R}$ be $(\eta_0, \theta_0, 0)$ with $\eta_0 > \bar{\eta}$.
The local direction of the coordinate lines is specified by the
vectors
\begin{align}
\hat{\bm{\eta}} &= \frac{1}{a}\, \frac{\cosh\eta \sin\theta\, \hat{\bm{\rho}}
                + \sinh\eta \cos\theta\, \hat{\bm{z}}}{\cosh^2\eta - \cos^2\theta}\,,
\\
\hat{\bm{\theta}} &= \frac{1}{a}\, \frac{\sinh\eta \cos\theta\, \hat{\bm{\rho}}
                - \cosh\eta \sin\theta\, \hat{\bm{z}}}{\cosh^2\eta - \cos^2\theta}\,,
\\                
\hat{\bm{\phi}} &= \frac{1}{a}\, \frac{\hat{\bm{\psi}}}{\sinh\eta \sin\theta}\,,
\quad a = \sqrt{a_z^2 - a_\perp^2}\,,
\end{align}
where $\hat{\bm{\rho}}$, $\hat{\bm{z}}$, and $\hat{\bm{\psi}}$ are the unit vectors in the radial, $z$, and the azimuthal directions, respectively.
Suppose the dipole moment $\mathbf{d}$ is also in the $x$--$z$ plane,
then it can be defined in terms of two coefficients, $d_\eta$ and $d_\theta$, such that
\begin{equation}
d_\eta = (\hat{\bm{\eta}} \textbf{d})\,,
\quad
d_\theta = (\hat{\bm{\theta}} \textbf{d})\,,
\end{equation}
where $\hat{\bm{\eta}}$ and $\hat{\bm{\theta}}$ are to be evaluated at $\eta = \eta_0$ and $\theta = \theta_0$.

As in the main text, we denote by $\Phi_1(\mathbf{r})$ and $\Phi_2(\mathbf{r})$ the scalar potentials inside and outside the spheroid, respectively.
We denote by $\Phi_d(\mathbf{r})$ the potential of the dipole alone.
These three potentials admit the expansions in series of spheroidal harmonics
$Y^m_l(\theta, \phi)$:
\begin{align}
Y^m_l &= \mathsf{P}^m_l(\cos\theta) \cos m \phi\,,
\\
\Phi_1 &=
\sum\limits_{l \,=\, 0}^{\infty} \sum\limits_{m \,=\, 0}^{l}
 D_l^m t_l^m \mathsf{P}^m_l(\cos\xi)
  Y^m_l
 \,,  \quad \xi < {\overline{\xi}}\,,
\label{eqn:Phi_1cos}\\
\Phi_2 &= \sum\limits_{l \,=\, 0}^{\infty} \sum\limits_{m \,=\, 0}^{l}
 D_l^m r_l^m Q^m_l(\cosh\eta)
 Y^m_l + \Phi_d
 \,,  \quad \eta > \bar\eta \,,
\label{eqn:Phi_2cos}\\
\Phi_d &=
 \sum\limits_{l \,=\, 0}^{\infty} \sum\limits_{m \,=\, 0}^{l}
 D_l^m \mathsf{P}^m_l(\cosh\eta)
  Y^m_l\,,
 \quad \eta < \eta_0\,.
\label{eqn:Phi_p}
\end{align}
The expansion coefficients $D_l^m$ of $\Phi_d$ can be derived from
the known expansion coefficients~\cite{Morse} $C_l^m$ of the potential of a point charge:
\begin{widetext}
\begin{align}
D_l^m &= \mathbf{d}\, \partial_{\mathbf{R}} C_{l m}
= \left( d_\eta \partial_{\eta_0}  + d_\theta \partial_{\theta_0} \right) C_l^m    \,,  \\
C_l^m (\mathbf{R}) &= \frac{\epsilon_m}{a}\,
 {i}^m (2l + 1) \left[
 \frac{(l - m)!}{(l + m)!}
 \right]^2
 Q_l^m(\cosh{\eta_0})\,
  \mathsf{P}_l^m(\cos{\theta_0}) \,,
\end{align}
where $\epsilon_m$ is the Neumann factor: $\epsilon_0 = 1$, $\epsilon_m = 2$ ($m = 1, 2, 3,\ldots$).
We obtain
\begin{equation}
D_l^m = \frac{\epsilon_m}{a}\,
 i^{m} (2l + 1) \left[
 \frac{(l - m)!}{(l + m)!}
  \right]^{2}
\left[d_\eta {\partial_{\eta_0}} Q_l^m(\cosh{\eta_0}) \mathsf{P}_l^m(\cos{\theta_0})
+  d_\theta Q_l^m(\cosh{\eta_0} ) {\partial_{\theta_0}} \mathsf{P}_l^m(\cos{\theta_0})
\right] \,.
\end{equation}
Imposing the boundary condition~\eqref{eqn:bc},
we get the matrix equation for the series coefficients:
\begin{equation}
\begin{pmatrix}
\mathsf{P}_l^m(\cos{\overline{\xi}}) & -Q_l^m(\cosh{\bar\eta}) \\
 i \sqrt{\varepsilon_{\perp}}  \sqrt{\varepsilon_{z}}\, \partial_{\overline{\xi}}  P_l^m(\cos{\overline{\xi}})
 & - \partial_{\bar\eta}  Q_l^m(\cosh{\bar\eta})
\end{pmatrix}
\begin{pmatrix}
t_l^m \\
r_l^m
\end{pmatrix}
\equiv \mathbf{M}\,
\begin{pmatrix}
t_l^m \\
r_l^m
\end{pmatrix}
=
\begin{pmatrix}
\mathsf{P}_l^m(\cosh{\bar\eta}) \\
\partial_{\bar\eta}\mathsf{P}_l^m(\cosh{\bar\eta})
\end{pmatrix}
\,,
\label{eqn:MA_DP}
\end{equation}
which has the solution
\begin{align}
t_l^m &= \frac{1}{\mathrm{det}\, \textbf{M}}\,
[-\partial_{\bar\eta}  Q_l^m(\cosh{\bar\eta})\, \mathsf{P}_l^m(\cosh{\bar\eta})\,
+ Q_l^m(\cosh{\bar\eta})\,   \partial_{\bar\eta} \mathsf{P}_l^m(\cosh{\bar\eta})]\,,
\label{eqn:A_1}\\
r_l^m &= \frac{1}{\mathrm{det}\, \textbf{M}}\,
S_l^m\,,
\label{eqn:A_2}   
\end{align}
where
\begin{align}
\mathrm{det}\, \textbf{M} &= -\mathsf{P}_l^m(\cos{\overline{\xi}}) \partial_{\bar\eta}\,  Q_l^m(\cosh{\bar\eta}) +
i\sqrt{\varepsilon_{\perp}} \sqrt{\varepsilon_{z}}\, \partial_{\overline{\xi}}  \mathsf{P}_l^m(\cos{\overline{\xi}})\, Q_l^m(\cosh{\bar\eta})
\label{eqn:detM}  \,,
\\
S_l^m &= -i \sqrt{\varepsilon_{\perp}} \sqrt{\varepsilon_{z}}\, \partial_{\overline{\xi}}  \mathsf{P}_l^m(\cos{\overline{\xi}})\, \mathsf{P}_l^m(\cosh{\bar\eta})
+ \mathsf{P}_l^m(\cos{\overline{\xi}})\,  \partial_{\bar\eta}\mathsf{P}_l^m(\cosh{\bar\eta})
\label{eq:S}
\,.
\end{align}
When eq~\eqref{eqn:exact} is satisfied,
$\mathrm{det}\, \textbf{M}$ vanishes, so that $t_l^m$ and $r_l^m$ diverge.
This behavior is consistent with having a divergent resonant response
at the polariton eigenfrequencies.
To compute Purcell's factor (Figure~5 of the main text and Figure~\ref{fig:radiative_damping}) we need
to know the response electric field outside the spheroid.
This field is given by
\begin{align}
\mathbf{E}_2^r &= -\partial_{\mathbf{r}}
 [\Phi_2(\mathbf{r}) - \Phi_d(\mathbf{r})]
= E_{2\eta}^r \hat{\bm{\eta}} + E_{2\theta}^r \hat{\bm{\theta}} + E_{2\phi}^r \hat{\bm{\phi}}\,,
\\
E_{2 i}^r &= -\sum\limits_{l \,=\, 0}^{\infty} \sum\limits_{m \,=\, 0}^{l}  D_l^m r_l^m
{\partial_i} \left[Q_{l}^{m} (\cosh{\eta})
 Y_l^m(\theta, \phi) \right]\,,
\quad
i = \eta, \theta, \phi\,,
\end{align}
so that Purcell's factor is
\begin{align}
 f = 1 + \frac{3}{2} \left(\frac{c}{\omega} \right)^3 \frac{1}{d^2} \, \mathrm{Im} \left[d_\eta E_{2\eta}^r + 
d_\theta E_{2\theta}^r \right]  \,.
\end{align}
Note also that the square of the response electric field is given by
\begin{align}
|\mathbf{E}_2^r|^2 &= \frac{1}{a^2} \left[
 \frac{(E_{2\eta}^r)^2}{\cosh^2\eta - \cos^2\theta}
  + \frac{(E_{2\theta}^r)^2}{\cosh^2\eta - \cos^2\theta}
  + \frac{(E_{2\phi}^r)^2}{\sinh^2\eta \sin^2\theta}
 \right]\,.
\end{align}  
\end{widetext}
The formula for the \textit{total} field $\mathbf{E}_1 = -\partial_{\mathbf{r}} \Phi_1(\mathbf{r})$ inside the spheroid is similar,
except it involves coefficients $t_l^m$.

The distribution of the electric field calculated at several periodic orbit frequencies of the bulk waves and the HSM are shown in Figure~6 of the main text.
They demonstrate an enhanced amplitude at the locations of the
classical trajectories launched from the point on a surface facing the dipole source. 
However, at a frequency away from the periodic orbit frequencies, wavepackets follow trajectories that spread all over the spheroid, forming an irregular background.
This effect is most apparent if we neglect the damping loss of the media, as shown in Figure \ref{fig:field_bulk}(a)--(f).
If the frequency is detuned by roughly $5\,\mathrm{cm}^{-1}$, i.e., a mere $0.3\%$ to either side off the periodic orbit frequency, the ray patterns disappear. The ray pattern of several surface periodic orbits are also found at their frequencies, as shown in panels $S_1$--$S_3$ of Figure~\ref{fig:field_bulk}.
If we account for the hBN phonon damping,
as we do in Figure~6 of the main text,
then the polariton propagation length becomes finite.
Notably, because of the scale-invariance of the problem, this length is not fixed, it scales in proportion to the size of the nanogranule.
This unusual property holds as long as the size of the granule is smaller than $c / \omega$, so that the scale-invariant quasi-static approximation is valid.
Therefore, a better measure of damping may not be the propagation length but rather the quality factor $Q = \omega / \Gamma$.
In Figure~6 of the main text we used 
the damping rate $\Gamma=7 \unit{cm^{-1}}$,
which is near the upper end of the experimentally determined range.\cite{Caldwell2014}
This corresponds to $Q \sim 200$.
Clearly, for such $\Gamma$ the polaritons still propagate far enough to complete the periodic orbits.
Additionally, the frequency windows for observing these orbits becomes wider, similar to the effect of damping on Purcell's factor resonances.

\subsection{Radiative correction}

To explain our procedure for computing the radiative damping,
it is instructive to consider two auxiliary problems first.
We begin with the textbook problem of a point-dipole emitter.
It is well known \cite{Ford.1984} that there is a small correction to the near field of such a dipole if we consider the retardation effect.
The correction contains a real part, which leads to a shift of the resonance frequency, and an imaginary part, which causes broadening of the linewidth and accounts for the energy loss due to radiation. We are primarily interested in the radiative damping; thus,
we retain only the imaginary part:
\begin{equation}
\mathbf{E}_d = \mathbf{E}_\mathrm{static} + i\, \mathrm{Im}\, \mathbf{E}_\mathrm{rad}
=-\frac{\mathbf{P} - 3(\hat{\mathbf{r}} \mathbf{P})\hat{\mathbf{r}}}{r^3}
 + \frac{2i}{3} k^3_0 \mathbf{P}.
\label{eqn:E_d}
\end{equation}
Next, consider a finite-size nanogranule subject to an external electric field. The field outside is the sum of incident field and dipole field. They have to satisfy boundary condition at the surface of the granule, which leads to 
\begin{align}
\mathbf{P} &= \hat{\chi}_0 (\mathbf{E}_0 + i\, \mathrm{Im}\, \mathbf{E}_\mathrm{rad} ) 
=\hat{\chi}_0 \left( \mathbf{E}_0 + \frac{2i}{3} k^3_0 \mathbf{P} \right),
\label{eqn:selfconsistent}
\end{align} 
where  $\hat{\chi}_0$ is the polarization tensor.
Solving eq.~\eqref{eqn:selfconsistent}, we get the radiative damping corrected polarization tensor:
\begin{align}
\mathbf{P} = \dfrac{\hat{\chi}_0} { 1-i\frac{2}{3}  (\frac{\omega}{c})^3  \hat{\chi}_0 } \mathbf{E}_0\,.
\end{align}
Finally, let us consider our original problem of a spheroidal granule perturbed by a dipole source.
Here the radiation correction field should be computed using the total dipole moment of the system: the source dipole and the spheroid.
From the first term of eq~\eqref{eqn:Phi_2cos},
the induced dipole moment of the granule has $x$- and $z$-components
\begin{align}
P_\mathrm{ind}^z=\frac{a^2}{3} D_1^0 r_1^0\,,\quad  P_\mathrm{ind}^x=\frac{2a^2}{3} D_1^1 r_1^1 \,.
\end{align}
The potential $\Phi_\mathrm{rad}$ corresponding to the correction $i\,\mathrm{Im}\, \mathbf{E}_\mathrm{rad}$ can be written as
\begin{align}
\Phi_\mathrm{rad}&= \sum\limits_{m \,=\, 0, 1} (c^m D_1^m r_1^m +  c^m_s d^z ) \mathsf{P}^m_1(\cosh\eta)Y^m_1 \,,
\end{align}
where
\begin{align}
c^0 &=-\frac{2i}{9} \left( \frac{\omega}{c} \right)^3 a^3\,,
\quad c^1 = - 2 c^0\,,\\
c^0_s &= -\frac{2i}{3} \left(\frac{\omega}{c} \right)^3 a\,,
\quad c^1_s = -c^0_s\,.
\end{align}
Thus, the right-hand side of eq~\eqref{eqn:Phi_2cos} changes to
\begin{align}
\Phi_2 &=  \Phi_d  +  \Phi_\mathrm{rad} +   \sum\limits_{l \,=\, 0}^{\infty} \sum\limits_{m \,=\, 0}^{l}
D_l^m r_l^m Q^m_l(\cosh\eta)
Y^m_l    \,.
\end{align}
Imposing the boundary condition~\eqref{eqn:bc}, we obtain the equation for the modified reflection coefficients $r_l^m$ of the dipolar (i.e., $m = 0, 1$) modes
\begin{widetext}
\begin{equation}
\mathbf{M}\,
\begin{pmatrix}
t_l^m \\
r_l^m
\end{pmatrix}
=
\left(1 + c^m r_l^m + c^m_s \frac{d^m}{D_l^m} \right)
\begin{pmatrix}
\mathsf{P}_l^m(\cosh{\bar\eta}) \\
\partial_{\bar\eta}\mathsf{P}_l^m(\cosh{\bar\eta})
\end{pmatrix}    \,,
\end{equation}
which can be rewritten as
\begin{equation}
\left[\mathbf{M}-c^m 
\begin{pmatrix}
0 & \mathsf{P}_l^m(\cosh{\bar\eta}) \\
0 & \partial_{\bar\eta}\mathsf{P}_l^m(\cosh{\bar\eta}
\end{pmatrix}
\right]
\begin{pmatrix}
t_l^m \\
r_l^m
\end{pmatrix}
=
\left(1 + c^m_s \frac{d^m}{D_l^m} \right)
\begin{pmatrix}
\mathsf{P}_l^m(\cosh{\bar\eta}) \\
\partial_{\bar\eta}\mathsf{P}_l^m(\cosh{\bar\eta})
\end{pmatrix}    \,.
\end{equation}

The solution for the reflection coefficient is
\begin{equation}
r_l^m = \frac{S_l^m}
             {\mathrm{det}\, \textbf{M} - c^{m} S_l^m}\,
\left(1 + c^m_s \frac{d^m}{D_l^m} \right) \,,
\label{eqn:reflection_corrected}
\end{equation}
where $\mathrm{det}\, \textbf{M}$ and $S_l^m$ can be found from eqs~\eqref{eqn:detM} and \eqref{eq:S}.
If $a \omega / c \ll 1$, i.e.,
if the nanogranule is much smaller than the diameter of Wheeler's radian sphere $c / \omega$, then $c^m, c^m_s \ll 1$, and the radiative damping is weak.
Far enough from the resonances, where $r_1^m$ is finite,
the correction to reflected field to the lowest order in $c^m, c^m_s$ is 
\begin{align}
\delta\Phi_2 &= \sum\limits_{m \,=\, 0, 1}
D_1^m \delta r_1^m Q^m_1(\cosh\eta)
Y^m_1 +  c^m D_1^m r_1^m  \mathsf{P}^m_1(\cosh\eta)Y^m_1   \\
&=\sum\limits_{m \,=\, 0, 1}
 (  c^m D_1^m r_1^m + c^m_s d^m) r_1^m Q^m_1(\cosh\eta)
Y^m_1 +  c^m D_1^m r_1^m  \mathsf{P}^m_1(\cosh\eta)Y^m_1 
\label{eq:radiative_correction}  \,.
\end{align}
\end{widetext}
The first term in eq~\eqref{eq:radiative_correction} is due to radiative correction of the polarizability of the spheroid, the second term is the reflected damping field of the source dipole, and the third term is the damping field of the induced dipole on the spheroid.
The corresponding change to Purcell's factor is
\begin{align}
\delta f = -\frac{3}{2} \left(\frac{c}{\omega} \right)^3 \frac{1}{d^2}\, \mathrm{Im}\left(d_\eta \partial_\eta \delta \Phi + 
d_\theta \partial_\theta \delta \Phi \right),
\label{eqn:delta_f_approx}
\end{align}
which is meant to be evaluated at $\eta = \eta_0$ and $\theta = \theta_0$.
In particular, if the dipole is polarized in the $x$-direction and is located on the positive-$z$ semi-axis,
then eq~\eqref{eqn:delta_f_approx} simplifies to
\begin{align}
	\delta f = \mathrm{Re} \left\{ r^1_1
	   \left[2\, \frac{Q^1_1}{P^1_1}
	   + r_1^1 \left(\frac{Q^1_1}{P^1_1} \right)^2
	   \right]
	\right\} \,,
\label{eqn:delta_f_example}
\end{align}
where the argument of both $Q^1_1$ and $P^1_1$ is equal to $\cosh\eta_0$.
Note that
$\delta f$ is free of the small parameter $a \omega / c$.
Hence, the ratio of the radiative correction and the uncorrected Purcell's factor scales linearly with $(a \omega / c)^3 \ll 1$.

\begin{figure}
\includegraphics[width=3.5 in]{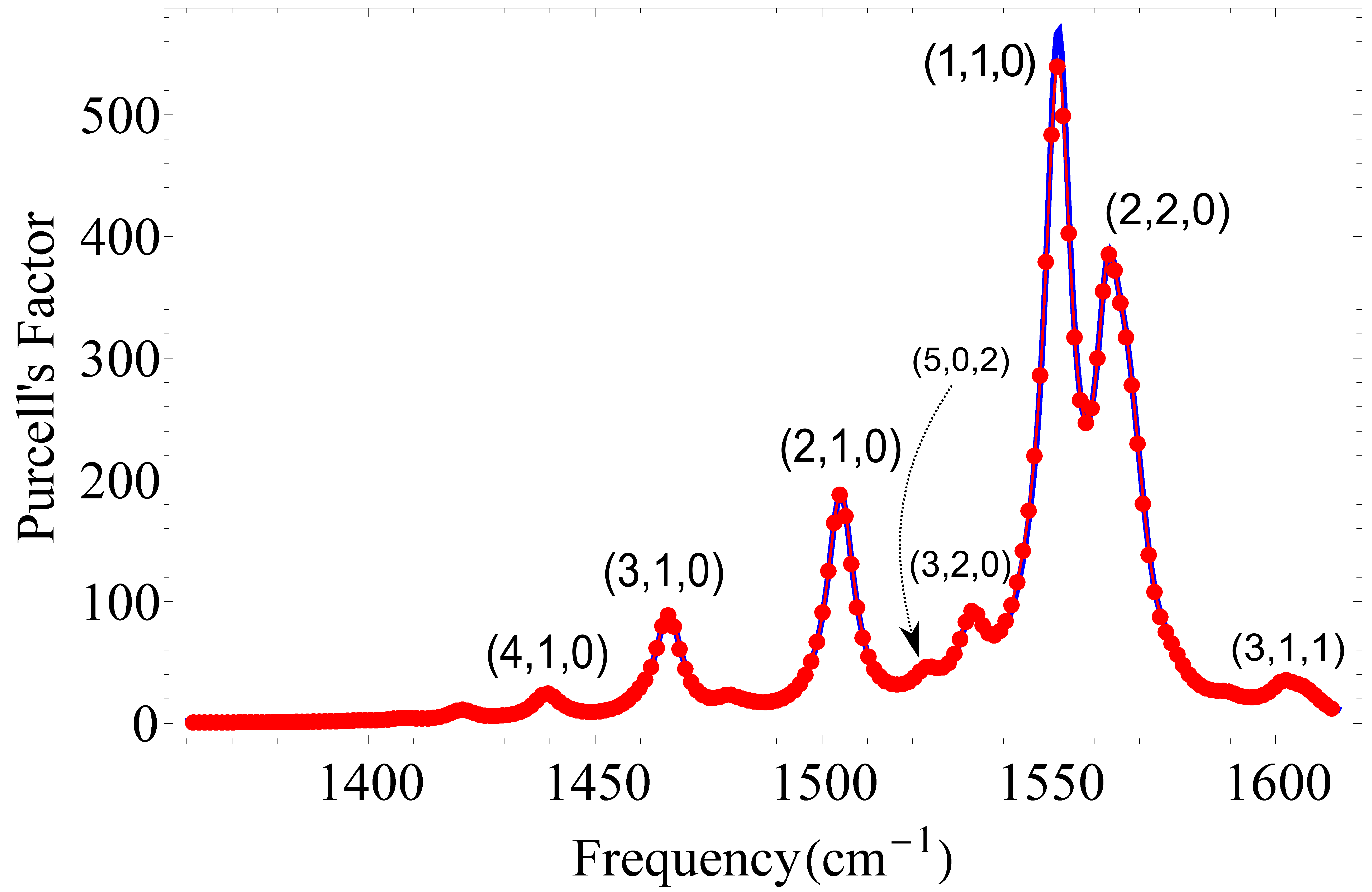}
	\caption{Purcell's factor as a function of frequency. The system configuration is the same as in Figure~5 of the main text except the long semi-axis of the spheroid is reduced to $200\unit{nm}$. The red dots (blue lines) show Purcell's factor with (without) the radiative damping included.}
	\label{fig:radiative_damping}
\end{figure}

At the resonances, $r_1^m$ diverges,
so one should avoid expanding the denominator in eq~\eqref{eqn:reflection_corrected}.
This leads to the more accurate formula
\begin{align}
\delta f = \mathrm{Re} \left\{
\frac{r_1^1}{1 - c^1 r_1^1}
\left[c^1 + 2\, \frac{Q^1_1}{P^1_1}
	   + r_1^1 \left(\frac{Q^1_1}{P^1_1} \right)^2
	   \right]
\right\}
\end{align}
instead of eq~\eqref{eqn:delta_f_example}.
This equation indicates that the resonances of the radiation-corrected Purcell's factor are shifted from points $r_1^1 = \infty$ to points $r_1^1 = 1 / c^1 \gg 1$.
Existence of such a shift is expected on general grounds.
However, we do not attempt to evaluate it
because a more important contribution to 
this shift should come from the real part of the radiative reaction field neglected in eq~\eqref{eqn:E_d}.

Calculations done according to the above formulas are shown in
Figure 5 of the main text for the case of spheroid with the long semi-axis $a_z = 500\unit{nm}$ and in Figure~\ref{fig:radiative_damping} for $a_z = 200\unit{nm}$.
The most noticeable effect in these Figures is the broadening of the resonance peaks.
There are two sources of such broadening.
The first is the intrinsic loss of the medium, described by the phonon damping rate $\Gamma$.
It influences all the modes, i.e.,
we can see that all the resonant peaks become broader as we increase $\Gamma$.
The second is the radiative damping effect, which broadens the $m = 0$ and $1$ dipolar modes but does not change much the linewidths of the remaining $(2, 2, 0)$ and $(3, 2, 0)$ modes.
However, for $a_z = 200\unit{nm}$ spheroid, $c^1$ is already so small that the radiative damping correction is negligible,
cf.~Fig.~\ref{fig:radiative_damping}.

\section{Dielectric function of \lowercase{h}BN}
Although our theory is developed for a general hyperbolic material, all the figures are calculated for the polar insulator hBN.
The dielectric tensor components of hBN have the following form:
\begin{align}
	\varepsilon_i (\omega)=\varepsilon_i (\infty) \left[ 
	1 + \frac{(\omega_i^{\mathrm{LO}})^2-(\omega_i^{\mathrm{TO}})^2}
	{(\omega_i^{\mathrm{TO}})^2 - \omega^2 - i \omega \Gamma_i}
	\right] \,,
	\label{eqn:dielectric}
\end{align}
where $i = \,\perp$ or $z$ and~\cite{Caldwell2014}
\begin{align}
\omega_\perp^\mathrm{TO} &=1360 \unit{cm^{-1}}, &\omega_\perp^\mathrm{LO} &=1614 \unit{cm^{-1}}\,,\\
\omega_z^{\mathrm{TO}} &=760 \unit{cm^{-1}}, &\omega_z^{\mathrm{LO}} &= 825 \unit{cm^{-1}}\,,\\
\varepsilon_\mathrm{\perp} (\infty) &= 4.90\,, &\varepsilon_{z} (\infty) &= 2.95 \,.
\end{align}
The results shown in Figure 5 of the main text are calculated for two values of the damping rate $\Gamma \equiv \Gamma_z$, namely, $4\unit{cm^{-1}}$ and $7\unit{cm^{-1}}$ to illustrate the effect of dielectric losses. 

We treated the nanogranule as a continuum medium.
As observed in recent experiments~\cite{Dai2014}
even for hBN as thin as three atomic layers
the continuum medium treatment that uses the bulk dielectric tensor yields an excellent agreement with the observed mode spectra.
Validity of the continuum medium treatment for few-layer hBN can also be justified theoretically based on the phonon dispersions of a few-layer hBN calculated by diagonalization of the full dynamical matrix.~\cite{Michel2011}
One can easily check that these results match with the continuum medium treatment for phonon momenta much smaller than the inverse lattice constant, which are relevant for our consideration.
The main difference between the continuum-medium and microscopic theories is the total number of phonon-polariton modes.
This number is finite and is proportional to the total number of layers.~\cite{Michel2011}
We have in mind nanogranules which contain hundreds or even thousands of layers.
For such granules, our continuum-medium theory should be fully valid. 

\bibliography{./library}

\end{document}